\documentclass[%
 reprint,
 amsmath,amssymb,
 aps,
 prb,
]{revtex4-1}

\usepackage{graphicx}
\usepackage{dcolumn}
\usepackage{bm}
\usepackage{mathrsfs}

\begin{document}

\preprint{APS/123-QED}

\title{Dynamics of the Geometric Phase in Inhomogeneous Quantum Spin Chains}

\author{Kaiyuan Cao}
 \email{191001004@njnu.edu.cn}
 \affiliation{Research Center for Intelligent Supercomputing, Zhejiang Laboratory, Hangzhou 311100, P. R. China}

\author{Shuxiang Yang}
 \affiliation{Research Center for Intelligent Supercomputing, Zhejiang Laboratory, Hangzhou 311100, P. R. China}

\author{Yayun Hu}
 \email{hyy@zhejianglab.com}
 \affiliation{Research Center for Intelligent Supercomputing, Zhejiang Laboratory, Hangzhou 311100, P. R. China}

\author{Guangwen Yang}%
 \email{ygw@tsinghua.edu.cn}
 \affiliation{Research Center for Intelligent Supercomputing, Zhejiang Laboratory, Hangzhou 311100, P. R. China}
 \affiliation{Department of Computer Science and Technology, Tsinghua University, Beijing, Haidian, P. R. China}%

\date{\today}

\begin{abstract}
  The dynamics of the geometric phase are studied in inhomogeneous quantum spin chains after a quench. Analytic expressions of the Pancharatnam geometric phase (PGP) $\mathcal{G}(t)$ are derived, for both the period-two quantum Ising chain (QIC) and the disordered QIC. In the period-two QIC, due to the periodic modulation, the PGP changes with time at the boundary of the Brillouin zone, and consequently, the winding number $\nu_{D}(t)=\int_{0}^{\pi}[\partial\phi_{k}^{G}(t)/\partial k]dk/2\pi$ based on the PGP is not quantized and thus not topological anymore. Nevertheless, the PGP and its winding number show non-analytic singularities at the critical times of the dynamical quantum phase transitions (DQPTs). This relation between the PGP and the DQPT is further confirmed in the disordered QIC, where the winding number is not defined. It is found that the critical time of DQPT inherited from the homogeneous system and the additional one induced by the weak disorder are also accompanied by the non-analytic singularity of the PGP, by decomposing the PGP into each quasiparticle mode. The connection between the non-analytic behavior of the PGP at the critical time and the DQPT, regardless of whether the winding number is topological, can be explained by the fact that they both arise when the Loschmidt amplitude vanishes.
\end{abstract}

\maketitle


\section{Introduction}

The geometric phase has seen remarkable advancement \cite{Xiao201082,Cooper201991,Bergholtz202193} since Berry published his seminal paper \cite{Berry1984392}, in which a quantum system is subjected to an adiabatically changing environment and manifests a geometric behavior in its phase \cite{Simon198351}. Later Aharanov and Anandan generalized the concept of Berry's phase to the cyclic evolution of the quantum system \cite{Aharonov198758}. In fact, it has been pointed out that Berry's phase appears in more general context, neither unitary nor cyclic, which is known as the Pancharatnam geometric phase (PGP) \cite{Samuel198860}. The geometric phase encodes the state of the system, and has been associated with a variety of condensed matter phenomena, such as the quantum Hall effect \cite{Thouless198249} and quantum phase transitions \cite{Carollo2020838,Zhang2015115}, etc.

Recently, the PGP has been proposed to characterize the dynamical quantum phase transition (DQPT) \cite{Budich201693}, which has attracted a lot of interest \cite{Sharma201693,Dutta201796,Vogel201714,Heyl201796,Bhattacharjee201897,
Lang201898,Qiu201898,Zhou201898,Mendl2019100,Qiu201920,Wang2019122,
Yang2019100,Zache2019122,Ding2020102,Xu20209,Zamani2020102,
Jafari2021103,Sadrzadeh2021103,Yu2021104,Zhou202133,Zhou202123,
Jafari2022105,Luan2022604,Naji2022105}. The DQPT describes the non-analytic behavior of the Loschmidt echo $\mathcal{L}(t)=|\mathcal{G}(t)|^{2}$ during the nonequilibrium dynamical evolution \cite{Heyl2013110,Zvyagin201642,Heyl201881}. The Loschmidt amplitude $\mathcal{G}(t)$ measures the overlap of the time-evolving state $|\psi(t)\rangle$ with its initial state $|\psi_{0}\rangle$, i.e.,
\begin{equation}\label{LA.definition}
  \mathcal{G}(t) = \langle\psi_{0}|\psi(t)\rangle = \langle\psi_{0}|e^{-iHt}|\psi_{0}\rangle,
\end{equation}
which is found formally analogical with the canonical partition function $Z(\beta)=\texttt{Tr}e^{-\beta H}$ of an equilibrium system. Similar to the equilibrium phase transition, the DQPTs can be seen from the cusp-like singularity of the rate function $\lambda(t)=-\lim_{N\rightarrow+\infty}\ln{[\mathcal{L}(t)]}/N$, which is also called the dynamical free energy density \cite{Karrasch201387}, and $N$ denotes the system size. Until now, the DQPT has been extensively studied in many theoretical \cite{Andraschko201489,Heyl2014113,Hickey201489,Vajna201489,Heyl2015115,
Schmitt201592,Vajna201591,Huang2016117,Bhattacharya201796,Bhattacharya201796014302,
Stamp201795,Halimeh201796,Homrighausen201796,Weidinger201796,Yang201796,
Kosior201897,Lang201898,Mera201897,Bojan2018120,Mehdi2019100,Huang2019122,
Lahiri201999,Liu201999,Cao2020102,Haldar2020101,Kyaw2020101,Wu2020101,
Halimeh2021104,Modak2021103,Cao202231,Jensen2022105,Weymann2022105,Hou2022106} and experimental \cite{Jurcevic2017119,Zhang2017551,Vogel201714,Guo201911,Wang2019122,Chen2020102,
Nie2020124,Tian2020124} works. Note that there exists another different definition of the DQPT, which studies the asymptotic
late-time steady state of the order parameter \cite{Yuzbashyan200696,Barmettler2009102,Eckstein2009103,Sciolla2010105,
Dziarmaga201059}. Two types of DQPTs have been found related in the long-range quantum Ising chain \cite{Bojan2018120}.

According to Berry's theory, a quantum system acquires a geometric phase $\phi^{G}(t)$ over the dynamical phase $\phi^{dyn}(t)$ during the time evolution \cite{Berry1984392}. The PGP \cite{Samuel198860} can be calculated by
\begin{equation}\label{definition.Pancharatnam.phase}
  \phi^{G}(t) = \phi(t) - \phi^{dyn}(t)
\end{equation}
with the total phase $\phi(t) = \texttt{arg}[\mathcal{G}(t)]$ and $\phi^{dyn}(t) = -\int_{0}^{t}ds\langle\psi(s)|H|\psi(s)\rangle$. One can define the winding number $\nu_{D}(t)=\int_{0}^{\pi}[\partial\phi_{k}^{G}(t)/\partial k]dk/2\pi$ as an integral of the momentum derivative of the PGP $\phi_{k}^{G}(t)$ over the Brillouin zone \cite{Budich201693}. A lot of works have shown that the winding number $\nu_{D}(t)$ is integer-quantized and changes by unit at the critical times of the DQPT \cite{Sharma201693,Dutta201796,Vogel201714,Heyl201796,Bhattacharjee201897,
Lang201898,Qiu201898,Zhou201898,Mendl2019100,Qiu201920,Wang2019122,
Yang2019100,Zache2019122,Xu20209,Zamani2020102,
Jafari2021103,Sadrzadeh2021103,Yu2021104,Zhou202133,Zhou202123,
Jafari2022105,Luan2022604,Naji2022105}, so that the winding number is treated as the \emph{dynamical topological order parameter} (DTOP) to characterize the DQPT. The PGP shows non-analytic singularities as dynamical vortices at the critical times of DQPTs \cite{Vogel201714,Mendl2019100}.  However, there exists clear evidence to show that the winding number $\nu_{D}(t)$ may be fractional-quantized and thus non-topological in the XY chain from a critical quantum quench \cite{Ding2020102}, although the DTOP is still one-to-one related to the DQPT. A significant question that follows is whether the DQPT and its associated PGP are not always accompanied by an integer-quantized (topological) winding number in general.

To answer this question, we investigate the PGP in two inhomogeneous systems: the period-two quantum Ising chain (QIC) and the disordered QIC. It is well-known that inhomogeneity can dramatically influence the behavior of DQPT \cite{Cao2020102,Cao202236}. The periodic modulation is found to induce richer DQPTs than those in the homogeneous system \cite{Cao202236}. New DQPTs appear after a quench across the critical lines of quantum phase transition under the influence of weak disorder \cite{Cao2020102}. Another ensuing interesting problem is whether the new extra DQPTs induced by the periodic modulation and the disorder are also related to the singularity of the PGP and its winding number. This is indeed the case in our work. The results reveal that the critical times of DQPTs induced by the periodic modulation and weak disorder can still be characterized by the non-analytic singularity of the PGP. However, the winding numbers $\nu_{D}(t)$ are found not to be quantized anymore in the period-two QIC. The reason for non-quantized winding numbers can be explained as that the PGP changes its values with time at the boundary of the Brillouin zone because of periodic modulation. This is different from the case in the XY chain from a critical quench \cite{Ding2020102}, where the fractional-quantized winding numbers are related to the singularity of the Bogoliubov angle at the gap-closing momentum. Moreover, the winding number is not properly defined in the disordered system due to the lack of translation symmetry, thus the DQPT induced by the weak disorder has no association with the topological change of the winding number. The results reflect that the winding number may not serve as the topological order parameter to characterize the DQPT in the general case. However, it is found that the DQPT and the singularity of PGP are closely connected in general, regardless of the quantization of the winding number. It can be understood from the fact that the vanishing of the Loschmidt amplitude contributes not only a cusp in the rate function for the DQPT and but also a dynamical vortex for the PGP.

The paper is organized in the following manner: in Sec.~\uppercase\expandafter{\romannumeral2}, we discuss the QIC with period-two nearest-neighbor interactions and give the formulas of the PGP and its winding number, more detailed derivations obtained in \textbf{Appendix.~A}; study the behavior of the PGP via two typical quench processes. In Sec.~\uppercase\expandafter{\romannumeral3}, we derive the PGP of the disordered QIC in real space for the first time, more detailed derivations obtained in \textbf{Appendix.~B}; similarly, give two typical examples to illustrate the behavior of the PGP in the disordered system. Finally, we summarize our results and draw the conclusion in Sec.~\uppercase\expandafter{\romannumeral4}.

\section{Periodic Quantum Spin Chains}

We consider the quantum Ising chain with the periodic nearest-neighbor interactions in the transverse field \cite{Tong200191,Tong200265,Titvinidze200332,Cao202236}. The Hamiltonian is given by
\begin{equation}\label{QIC.period}
  H = -\frac{1}{2}\sum_{n=1}^{N}J_{n}\sigma_{n}^{x}\sigma_{n+1}^{x}
      -\frac{h}{2}\sum_{n=1}^{N}\sigma_{n}^{z},
\end{equation}
where $\sigma^{a}(a= x,y,z)$ are the Pauli matrices, $J_{n}$ are the strength of interactions between the nearest-neighbor spins, and $h$ is the external transverse field, respectively. We consider the QIC with period-two nearest-neighbor interactions $(l\in\mathbb{Z})$
\begin{equation}\label{NN.interaction.two}
  J_{n}=\left\{\begin{array}{ll}
                 J, & n=2l-1, \\
                 J_{1}, & n=2l.
               \end{array}
  \right.
\end{equation}
For convenience, we set $\alpha=J_{1}/J$ and $J=1$ without losing generality. The period-two QIC undergoes the quantum phase transition from the ferromagnetic (FM) phase to the paramagnetic (PM) phase at the critical point $h_{c}=\sqrt{\alpha}$, when the external field $h$ increases \cite{Pfeuty1979245,Tong200191}.

We can solve the Hamiltonian (\ref{QIC.period}) via the Jordan-Wigner and Bogoliubov transformations (see \textbf{Appendix.~A.~1}), where the diagonal form of Hamiltonian is
\begin{equation}\label{diagonal.two}
  H = \sum_{k}\Lambda_{k1}(\eta_{k1}^{\dag}\eta_{k1}-\frac{1}{2}) + \Lambda_{k2}(\eta_{k2}^{\dag}\eta_{k2}-\frac{1}{2}).
\end{equation}
Unlike that in the homogeneous QIC, the period-two QIC has two quasiparticle excitation spectra $\Lambda_{k1}$ and $\Lambda_{k2}$. The zero-point (ground-state) energy of this spinless fermion system is given by
\begin{equation}\label{ground.state.energy}
  E_{0} = \sum_{k>0}E_{0k} = -\sum_{k>0}(\Lambda_{k1}+\Lambda_{k2}),
\end{equation}
and the ground state is $|GS\rangle=\bigotimes_{k>0}|GS_{k}\rangle$ with $|GS_{k}\rangle = |0_{k1}0_{-k1}0_{k2}0_{-k2}\rangle$ for every $k$ $(k>0)$.

We study the nonequilibrium dynamical evolution induced by a quantum quench. The system is prepared in the ground state $|\psi_{0}\rangle = \bigotimes_{k>0}|\psi_{0k}\rangle, |\psi_{0k}\rangle=|GS_{k}\rangle$ of an initial Hamiltonian $H_{0} = H(h_{0})$. At time $t = 0$, the external field will be changed suddenly to another value $h_{1}$, that corresponds to the Hamiltonian $\tilde{H} = H(h_{1})$. In this section, we use $\tilde{\eta}_{k}(\tilde{\eta}^{\dag}_{k})$, $|\tilde{\psi}_{0k}\rangle$, and $\tilde{\Lambda}_{k}$ to denote the corresponding items of the post-quench Hamiltonian $\tilde{H}$. The time-evolved state is given by
\begin{equation}\label{time.state.two}
  |\psi_{k}(t)\rangle = e^{-i\tilde{H}t}|\psi_{0k}\rangle.
\end{equation}
By decomposing the Loschmidt amplitude $\mathcal{G}(t) = \prod_{k>0}\mathcal{G}_{k}(t)$, we obtain
\begin{equation}\label{LA.k.two}
  \mathcal{G}_{k}(t)  =\frac{e^{-i\tilde{E}_{0k}t}}{\mathcal{N}^{2}}\prod_{\mu,\nu=1}^{2}
  [1+|G_{k\mu,-k\nu}|^{2}e^{-i(\tilde{\Lambda}_{k\mu}+\tilde{\Lambda}_{-k\nu})t}],
\end{equation}
where $G = -(U\tilde{U}^{\dag}+V\tilde{V}^{\dag})^{-1}(U\tilde{V}^{T}+V\tilde{U}^{T})$ is an antisymmetric matrix dependent on the parameters of the pre- and post-quench Hamiltonian (see \textbf{Appendix.~A.~2}). Similar to the Lee-Yang zeros, we can illustrate the DQPT in a straightforward way via the Fisher zeros in the complex time plane \cite{Heyl2013110}. From $\mathcal{G}_{k}(z)=0, \texttt{Im}(z)=t$, the Fisher zeros of the Loschmidt amplitude for every $k$ are given by
\begin{equation}\label{Fisherzero.two}
  z_{n}(k, \mu, \nu)=\frac{1}{\tilde{\Lambda}_{k\mu}+\tilde{\Lambda}_{-k\nu}}[\ln{|G
  _{k\mu,-k\nu}|^{2}}+i(2n+1)\pi]
\end{equation}
with $\mu,\nu = 1,2$. The Fisher zeros will have an intersection with imaginary axis of the complex time plane when the DQPT occurs. Eq.~(\ref{Fisherzero.two}) implies that Fisher zeros in the period-two QIC have multiple branches which are different from the single branch in the homogeneous QIC\cite{Cao202236}. With the help of the Fisher zeros (\ref{Fisherzero.two}), we can easily obtain the critical momentum $k_{c}$ of the DQPT which satisfies $|G_{k_{c}\mu,-k_{c}\nu}|=1$, and the associated critical time
\begin{equation}\label{critical.t.two}
  t_{c}(n) = \frac{(2n+1)\pi}{\tilde{\Lambda}_{k_{c}\mu}+\tilde{\Lambda}_{-k_{c}\nu}}.
\end{equation}

To study the behavior of the PGP, we rewrite the Loschmidt amplitude $\mathcal{G}_{k}(t)$ in the polar coordinate, which is
\begin{equation}\label{polar.loschmidt.amplitude.two}
  \mathcal{G}_{k}(t) = r_{k}(t)e^{i\phi_{k}(t)} = r_{k}(t)e^{i[\phi_{k}^{dyn}(t)+\phi_{k}^{G}(t)]},
\end{equation}
where $\phi_{k}^{dyn}(t)$ and $\phi_{k}^{G}(t)$ are dynamical phase and purely geometric phase, respectively. According to Eq.~(\ref{polar.loschmidt.amplitude.two}), we get the dynamical free energy (rate function) in the thermodynamic limit as
\begin{equation}\label{rate.function.two}
  \lambda(t) = -\int_{0}^{\pi}\frac{dk}{2\pi}\ln{r_{k}^{2}(t)}.
\end{equation}
Clearly, the rate function $\lambda(t)$ only depends on the modulus $r_{k}(t)$ of the Loschmidt amplitude $\mathcal{G}_{k}(t)$. However, at the critical momentum $k_{c}$, $\lambda(t)$ has a non-analytic point, i.e., $r_{k_{c}}(t)=0$. According to the basic theory in complex math, when a complex number has a zero modulation, its argument angle can take any value. This will be reflected by a dynamical vortex (non-analytic singularity) in the PGP \cite{Vogel201714,Mendl2019100}. This is the essential reason why DQPTs can be characterized by the PGP.

The PGP $\phi_{k}^{G}(t)$ can be extracted from the time-dependent argument $\phi_{k}(t)$ of the Loschmidt amplitude by
\begin{equation}\label{geometric.phase.two}
  \phi_{k}^{G}(t) = \phi_{k}(t) - \phi_{k}^{dyn}(t),
\end{equation}
where the dynamical phase $\phi_{k}^{dyn}(t)$ is
\begin{equation}\label{dyn.phase.two}
  \begin{split}
    \phi_{k}^{dyn}(t) & = -\int_{0}^{t}ds\langle\psi(s)|\tilde{H}_{k}|\psi_{s}\rangle
    \\
    & = \{[1-\frac{2(|G_{k1,-k1}|^{2}+|G_{k1,-k2}|^{2})}{(1+|G_{k1,-k1}|^{2})(1+|G_{k1,-k2}|^{2})}]
            \tilde{\Lambda}_{k1} \\
      & + [1-\frac{2(|G_{-k1,k2}|^{2}+|G_{k2,-k2}|^{2})}{(1+|G_{-k1,k2}|^{2})(1+|G_{k2,-k2}|^{2})}]
            \tilde{\Lambda}_{k2}\} t.
  \end{split}
\end{equation}
The dynamical phase $\phi_{k}^{dyn}(t)$, which is found proportion to time $t$, is always an analytic function. Therefore, the non-analytic behavior of the argument $\phi_{k}(t)$ will be reflected in the PGP $\phi_{k}^{G}(t)$ at the critical time, where $\phi_{k}^{G}(t)$ is ill-defined.

Note that the PGP $\phi_{k}^{G}(t)$ is usually folded into its principal angle value, i.e., $\phi_{k}^{G}(t)\in(-\pi,\pi]$. The associated winding number in terms of the PGP can be calculated by
\begin{equation}\label{winding.number.two}
  \nu_{D} = \frac{1}{2\pi}\int_{0}^{\pi}\frac{\partial\phi_{k}^{G}(t)}{\partial k}dk = \frac{\phi_{k=\pi}^{G}(t)-\phi_{k=0}^{G}(t)}{2\pi}+\mathscr{N},
\end{equation}
where $\mathscr{N}$ is the folding number of the PGP from $-\pi$ to $\pi$ or from $\pi$ to $-\pi$ when $\phi_{k}^{G}(t)$ exceeds its principal value interval. The folding number $\mathscr{N}$ minuses one when the PGP is folded from $-\pi$ to $\pi$, and pluses one when the PGP is folded from $\pi$ to $-\pi$.

In the literatures\cite{Sharma201693,Dutta201796,Vogel201714,Heyl201796,Bhattacharjee201897,
Lang201898,Qiu201898,Zhou201898,Mendl2019100,Qiu201920,Wang2019122,
Yang2019100,Zache2019122,Xu20209,Zamani2020102,
Jafari2021103,Sadrzadeh2021103,Yu2021104,Zhou202133,Zhou202123,
Jafari2022105,Luan2022604,Naji2022105}, $\phi_{k}^{G}(t)$ are found pinned to zero at the boundary of the Brillouin zone in the homogeneous systems, i.e., $\phi_{k=\pi}^{G}(t)-\phi_{k=0}^{G}(t)=0$. This ensures that the winding number $\nu_{D}(t)$ is integer-quantized. However,the situation is different in our periodic case. It is found that the PGP changes its value with time at the boundary of the Brillouin zone under the periodic modulation, which results in the winding number not quantized. In the following, we will show our interesting findings with two typical examples.

\subsection{Quench from the FM phase to the PM phase}

We further investigate the PGP $\phi_{k}^{G}(t)$ and the associative winding number $\nu_{D}(t)$ for the period-two QIC by showing two typical quench examples. We take the value $\alpha=J_{1}/J=0.5$, which implies the system undergoes an Ising transition at the critical point $h_{c}=\sqrt{\alpha}\approx0.707$.

\begin{figure}
  \centering
  \includegraphics[width=1.0\linewidth]{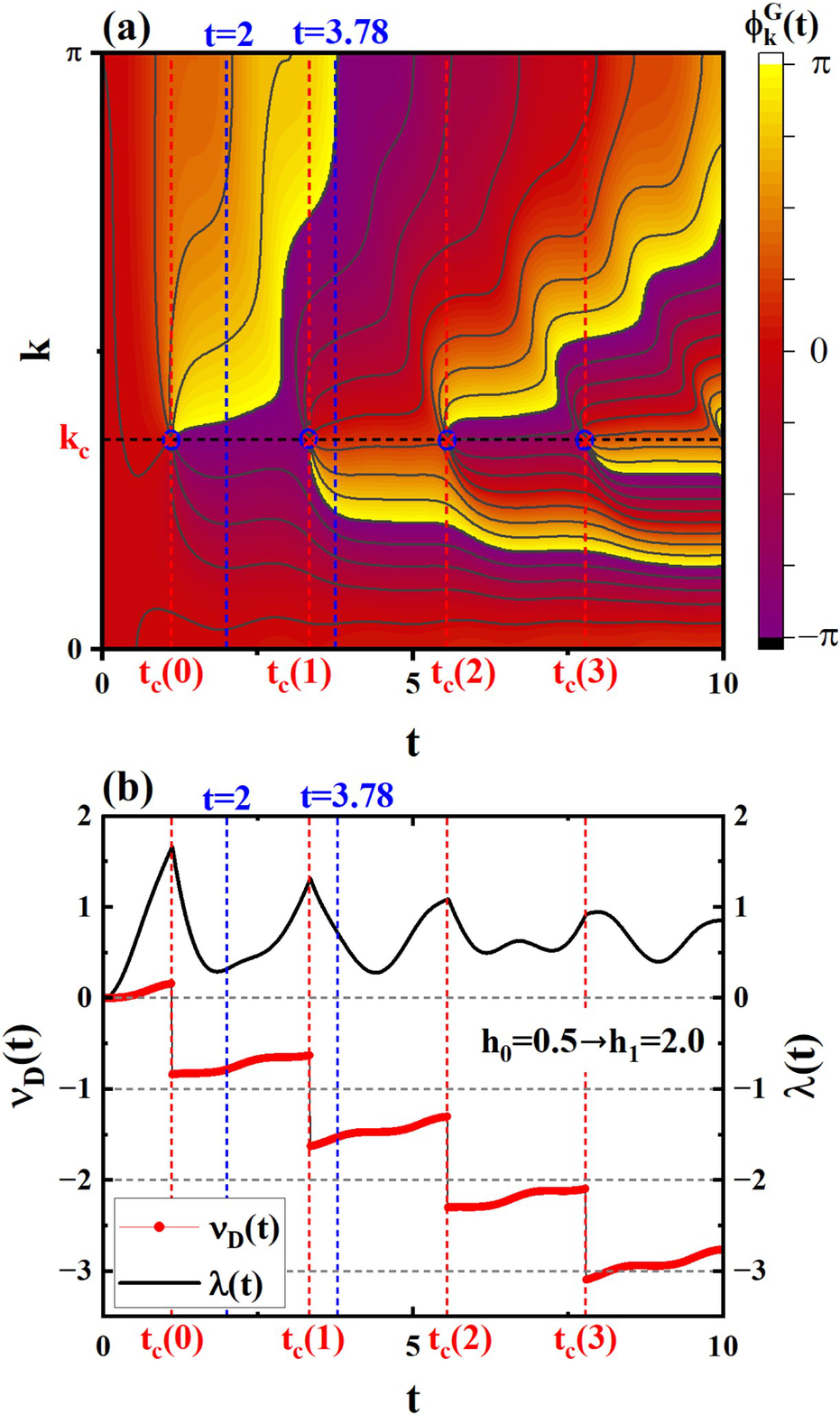}\\
  \caption{(color online) (a) Contour plot of PGP $\phi_{k}^{G}(t)$ as a function of $(k,t)$ for the quench from the FM phase to the PM phase ($h_{0}=0.5$ to $h_{1}=2.0$). The phase vortices are marked in blue circles at the critical momentum and critical times $(k_{c},t_{c}(n))$. The red ``$\times$'' denotes the critical momentums and critical times obtained according to Eqs.~(\ref{Fisherzero.two}) and (\ref{critical.t.two}).
  (b) The time evolutions of the winding number $\nu_{D}(t)$ (red scatter line) and the rate function $\lambda(t)$ (black line) are plotted in comparison. It can be seen that $\nu_{D}(t)$ is not integer-quantized, but jumps discontinuously at the critical times $t_{c}(n),n=0,1,2,3,\cdots$.}\label{topological.two.FMPM}
\end{figure}

\begin{figure}
  \centering
  \includegraphics[width=1.0\linewidth]{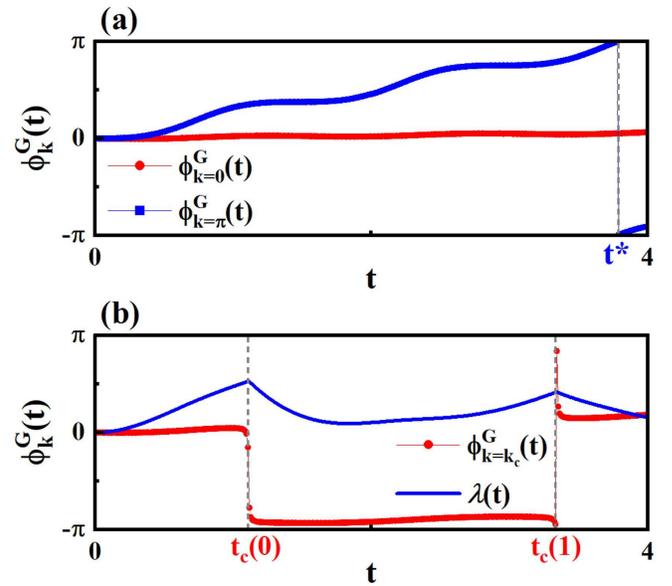}\\
  \caption{(color online) Factor components of the PGP are plotted as a function of $t$, (a) for $k=0$, (b) for $k=\pi$, and (c) for $k=k_{c}\approx1.10$, respectively. At the boundary of the Brillouin zone, (a) $\phi_{k=0}^{G}(t)\approx0$ is almost constant, but (b) $\phi_{k=\pi}^{G}(t)$ changes its value with time. Note that the jump of $\phi_{k=\pi}^{G}(t)$ at time $t^{*}$ results from restricting the PGP to its principal angle value, which will not lead to the presence of the DQPT. (c) While for the critical momentum $k_{c}$, $\phi_{k_{c}}^{G}(t)$ shows nonanalytic singularities at critical times $t=t_{c}(0)$ and $t_{c}(1)$. The black line is the rate function $\lambda(t)$ which shows singularities at critical times $t=t_{c}(0)$ and $t_{c}(1)$. }\label{Pancharatnam.period.FMPM.k}
\end{figure}

First, we study the case of quench from the FM phase to the PM phase.
In Fig.~\ref{topological.two.FMPM}~(a), we show the contour plot of PGP $\phi_{k}^{G}(t)$ as a function of $(k,t)$, where the quench path is from $h_{0}=0.5$ to $h_{1}=2.0$. The symmetry of the Hamiltonian and the initial state ensures $\phi_{k}^{G}(t)=\phi_{-k}^{G}(t)$, so throughout this paper we only show the PGP in the $(k,t)$ plane for $k>0$. We obtain the critical momentum $k_{c}$ and the critical times $t_{c}(n)$ of the DQPTs, which are marked by red ``$\times$'', according to Eqs.~(\ref{Fisherzero.two}) and (\ref{critical.t.two}). Obviously, there is one critical momentum $k_{c}\approx1.10$ corresponding to multiple critical times $t_{c}(n)=(2n+1)t_{c}(0),n=0,1,\cdots$ [see Fig.~\ref{topological.two.FMPM}~(a)]. It can be seen that the $\phi_{k}^{G}$ shows non-analytic singularities (dynamical vortices circled in blue) at the critical times $t_{c}(n)$ and critical momentum $k_{c}$.
Furthermore, we notice that $\phi_{k}^{G}(t)$ does not complete full circles in the Brillouin zone, i.e., $\phi_{k=\pi}^{G}(t)-\phi_{k=0}^{G}(t)\neq2n\pi$. For instance, when $t=2$, $\phi_{k}^{G}(t)$ changes its values by $0\rightarrow-\pi\stackrel{\text{folding}}{\longrightarrow}\pi\rightarrow0.46\pi$ [see the blue line $t=2$ in Fig.~\ref{topological.two.FMPM}~(a)]; when $t=3.78$, $\phi_{k}^{G}(t)$ changes its values by $0\rightarrow-\pi\stackrel{\text{folding}}{\longrightarrow}\pi\rightarrow-\pi\stackrel{\text{folding}}{\longrightarrow}\pi$ [see the blue line $t=3.78$ in Fig.~\ref{topological.two.FMPM}~(a)].
Here $-\pi\stackrel{\text{folding}}{\longrightarrow}\pi$ denotes restricting $\phi_{k}^{G}(t)$ to its principal angle value $(-\pi,\pi]$. This implies that the associated winding number $\nu_{D}(t)$ may not be an integer, according to Eq.~(\ref{winding.number.two}).
To establish that this is indeed the case, we calculate and plot the winding number $\nu_{D}(t)$ as a function of $t$ in Fig.~\ref{topological.two.FMPM}~(b). In order to see the critical times clearly, we also show the corresponding rate function $\lambda(t)$. It is clear that the winding number $\nu_{D}(t)$ is not integer-quantized. Specifically, when $t=2$, the winding number $\nu_{D}(t)\approx-0.77$ [see the blue line $t=2$ in Fig.~\ref{topological.two.FMPM}~(b)]; when $t=3.78$, $\nu_{D}(t)\approx-1.5$ [see the blue line $t=3.78$ in Fig.~\ref{topological.two.FMPM}~(b)]. Nevertheless, the winding number $\nu_{D}(t)$ is found to jump discontinuously at the critical times of the DQPTs. This means that the winding number can still be the detector of the DQPTs although it is not topological.

We now focus on three factor components of the PGP $\phi_{k}^{G}(t)$ for the momentum $k=0$, $k=\pi$, and $k=k_{c}$ [see Fig.~\ref{Pancharatnam.period.FMPM.k}~(a), (b), and (c)] respectively, which are closely relevant to $\nu_{D}(t)$ according to Eq.~(\ref{winding.number.two}). It is found that at the boundary of the Brillouin zone, $\phi_{k=0}^{G}(t)\approx0$ is almost constant all the time, but $\phi_{k=\pi}^{G}(t)$ changes its value with time. We have tested other parameters and find this to  be a general behavior in the case of quench from the FM phase to the PM phase.  Note that the jump of $\phi_{k=\pi}^{G}(t)$ at time $t^{*}$ results from restricting the PGP to its principal angle value, which will not lead to the presence of the DQPT. For the critical momentum $k_{c}\approx1.10$ of the DQPT, we can see that $\phi_{k_{c}}^{G}(t)$ has non-analytic points at times $t=t_{c}(0), t_{c}(1)$, which are exactly the critical times of the DQPT.

\begin{figure}[t]
  \centering
  \includegraphics[width=1.0\linewidth]{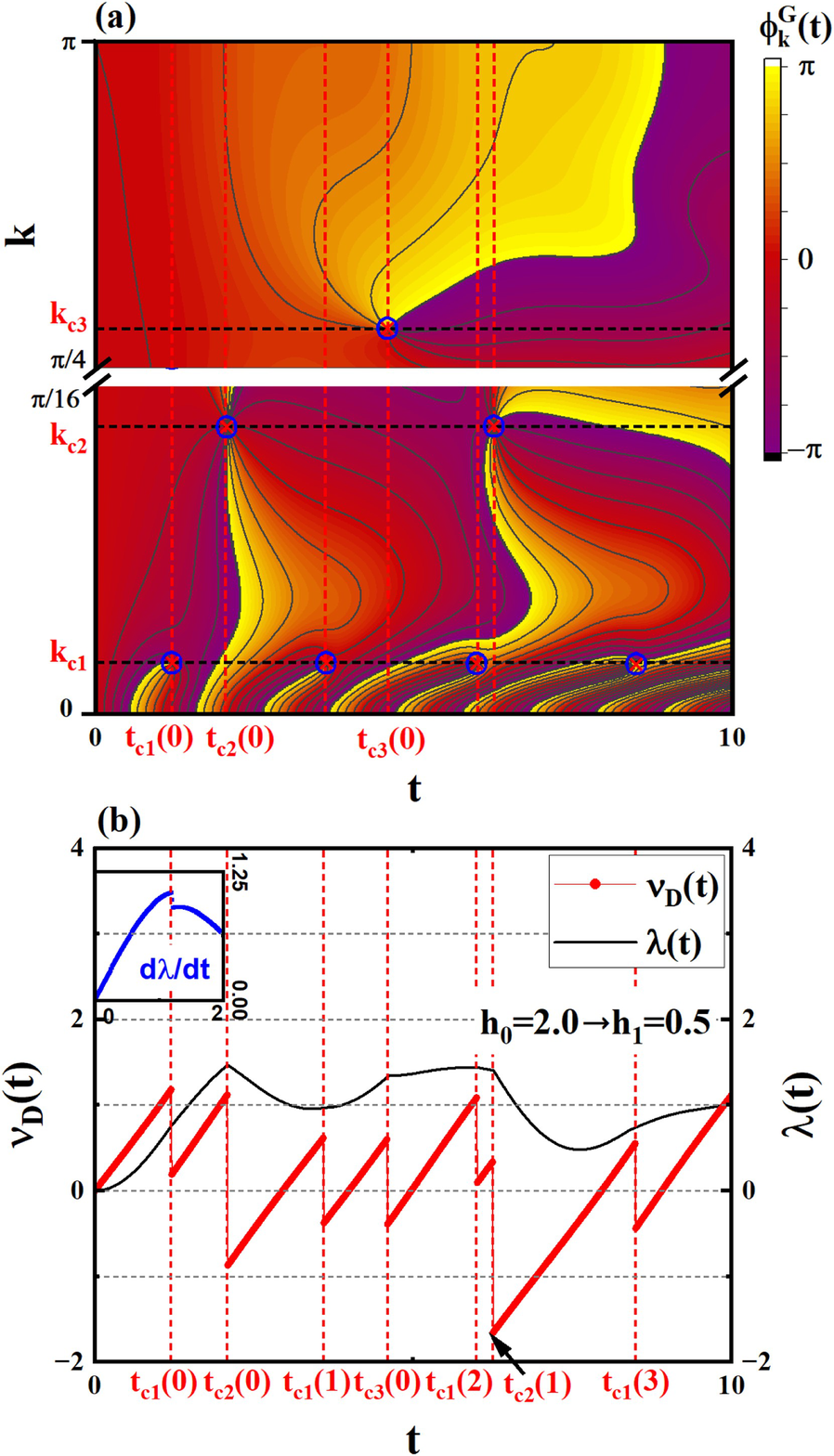}\\
  \caption{(a) Contour plot of PGP $\phi_{k}^{G}(t)$ as a function of $(k,t)$ for the quench from the PM phase to the FM phase ($h_{0}=2.0$ to $h_{1}=0.5$). The inset graph is shown to see the case for small $k$ clearly. The phase vortices are marked in blue circles at the critical momentum and critical times $(k_{c},t_{c}(n))$. The red ``$\times$'' denote the critical momentums and critical times obtained according to the Fisher zeros (\ref{Fisherzero.two}) and Eq.~(\ref{critical.t.two}). (b)The winding number $\nu_{D}(t)$ and rate function $\lambda(t)$ as functions of time $t$. Note that the critical time $t_{c1}(0)$ is not distinguished clearly, so that we use the first-order derivative $d\lambda/dt$ of the rate function to highlight the singularity [see the inset graph].}\label{topological.two.PMFM}
\end{figure}

\subsection{Quench from the PM phase to the FM phase}

As a second example, we consider the case of quench from the PM phase to the FM phase. In Fig.~\ref{topological.two.PMFM}~(a), we show the contour plot of the PGP $\phi_{k}^{G}(t)$ as a function of $(k,t)$. Here the quench path is from $h_{0}=2.0$ to $h_{1}=0.5$, which is the inverse path of the previous example shown in the Fig.~\ref{topological.two.FMPM} and \ref{Pancharatnam.period.FMPM.k}. Unlike that in the case of quench from the FM phase to the PM phase, there are three critical momentums $k_{c1}, k_{c2}$ and $k_{c3}$ corresponding to three groups of critical times $t_{c1}(n)=(2n+1)t_{c1}(0)$, $t_{c2}(n)=(2n+1)t_{c2}(0)$, and $t_{c3}(n)=(2n+1)t_{c3}(0), n=0,1,2,\cdots$. This can be understood based on Eq.~(\ref{Fisherzero.two}), i.e. three branches of Fisher zeros have intersections with the imaginary axis in the complex time plane \cite{Cao202236}. At the critical momentum and critical times $(k_{c},t_{cm}(n)), m=1,2,3$, $\phi_{k}^{G}(t)$ shows dynamical phase vortices circled in blue [see Fig.~\ref{topological.two.PMFM}~(a)]. Similar to the case of quench from the FM phase to the PM phase, $\phi_{k}^{G}(t)$ does not complete full circles in the Brillouin zone. For instance, when $t=0.53$, $\phi_{k}^{G}(t)$ changes its values by $\pi\stackrel{\texttt{folding}}{\longrightarrow}-\pi\rightarrow0$, which implies the corresponding winding number $\nu_{D}(t=0.53)\approx0.5$. We plot the winding number $\nu_{D}(t)$ and rate function $\lambda(t)$ in Fig.~\ref{topological.two.PMFM}~(b). It can be seen that the winding number $\nu_{D}(t)$ shows approximately linear change with time $t$ within two neighbouring critical times. As expected, the discontinuous points of $\nu_{D}(t)$ are accompanied by the critical times of DQPTs and the dynamical vortices in PGP [see Fig.~\ref{topological.two.PMFM}~(b)].

By the way, we also investigate the case of quench within the same phase in \textbf{Appendix.~A.~3} [see Fig.~\ref{topological.same.phase}], where the DQPT is absent. It is clear that the PGP $\phi_{k}^{G}(t)$ is analytic on the $(k,t)$ plane, and its winding number $\nu_{D}(t)$ is a continuous function of time when the DQPT does not occur. All the examples reveal that the PGP and the winding number are not topological in the periodic-two QIC which is different from that in homogeneous systems \cite{Budich201693,Sharma201693,Zhou201898,Qiu201920,Yang2019100,
Jafari2021103}. In both periodic and homogeneous systems, the discontinuous points of the winding number $\nu_{D}(t)$ and the dynamical vortices in the PGP are closely related to the critical times of DQPTs, and they occur when the Loschmidt amplitude equals zero.

\begin{figure}
  \centering
  \includegraphics[width=1.0\linewidth]{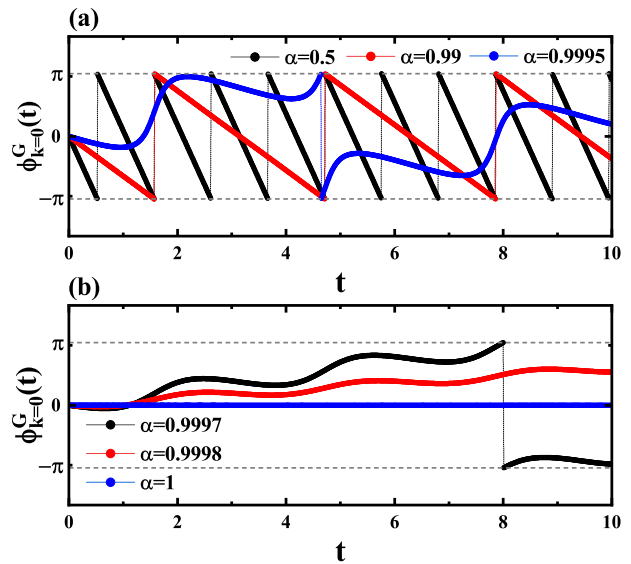}\\
  \caption{Factor components $\phi_{k=0}^{G}(t)$ for different parameters $\alpha$ in the case of quench from the PM phase to the FM phase ($h_{0}=2.0$ to $h_{1}=0.5$). (a) The speed of oscillation tends to decrease with $\alpha$ from $0.5$ to $1$. (b) In particular, the oscillation period goes to the infinity as $\alpha$ approaches unity, and $\phi_{k=0}^{G}(t)$ is zero for the homogeneous system ($\alpha=1$). }\label{period.to.homogeneous}
\end{figure}

In previous works where the homogeneous systems are intensively studied \cite{Budich201693,Sharma201693,Zhou201898,Qiu201920,Yang2019100,Jafari2021103}, the winding number $\nu_{D}(t)$ itself is integer-quantized, and so is the discontinuous jumps of $\nu_{D}(t)$ at the critical time of the DQPT. The quantized jumps of $\nu_{D}(t)$ at DQPTs are also observed in period-two QIC, but $\nu_{D}(t)$ is no longer quantized. The discrepancy of these two types of quantization can be traced back to different physical origins. The quantization of the jump is protected by the dynamical vortex of the PGP in the $(k,t)$ space. The PGP diverges at the dynamical vortex where the Loschmidt amplitude $\mathcal{G}(t) = r(t)e^{[i(\phi^{G}(t)+\phi^{dyn}(t)]}$ vanishes and its phase is ill-defined. However, according to Eq.~(\ref{winding.number.two}), both the boundary term and the jump term contribute to the winding number. Although the jump term always provides quantized contributions as just explained, the boundary term $[\phi_{k=\pi}^{G}(t)-\phi_{k=0}^{G}(t)]$ is not necessarily quantized in general. For example, in the period-two QIC, as the PGP changes value with time at the boundary of the Brillouin zone ($k=0$ and $k=\pi$) in the presence of periodic modulation.
To illustrate the effect of periodic modulation, we show the factor component $\phi_{k=0}^{G}(t)$ for different parameters $\alpha$ in Fig.~\ref{period.to.homogeneous}, where $\alpha=1$ corresponds to the homogeneous system. It is clear that $\phi_{k=0}^{G}(t)$ oscillates with time in the periodic QIC, and the speed of oscillation tends to decrease with $\alpha$ from $0.5$ to $1$ [see Fig.~\ref{period.to.homogeneous}~(a) and (b)]. In particular, $\phi_{k=0}^{G}(t)$ is zero in the homogeneous system ($\alpha=1$). Therefore, we conclude that the non-quantized winding number results from the periodic modulation. Actually, the change of the PGP with time at the boundary of the Brillouin zone is also observed in the periodic Kitaev chain \cite{Mendl2019100}. Therefore, it is inferred that the winding number in the periodic Kitaev chain is not quantized either.

\section{Disordered Quantum Spin Chains}

In this section, we extend the PGP to disordered systems. The Hamiltonian of the QIC with disordered hopping interactions is
\begin{equation}\label{disorder.Hamiltonian}
  H = -\frac{1}{2}\sum_{n=1}^{N}J_{n}\sigma_{n}^{x}\sigma_{n+1}^{x}-\frac{h}{2}\sum_{n=1}^{N}\sigma_{n}^{z},
\end{equation}
where $J_{n}=J+\Delta J_{n}$ are the hopping interactions between the nearest neighbor spins. $\Delta J_{n}$ are independent random numbers distributed uniformly in the interval $[-w/2,w/2]$ with the strength of disorder $w$. For convenience, we take $J=1$ without loss of generality.

By using the Jordan-Wigner and Bogoliubov transformations\cite{Lieb196116,Suzuki2013}, the Hamiltonian in Eq.~(\ref{disorder.Hamiltonian}) can be reduced to the diagonal form (\textbf{see Appendix.~B.~1})
\begin{equation}\label{disorder.diagonal.Hamiltonian}
  H = \sum_{n}\Lambda_{n}(\eta_{n}^{\dag}\eta_{n}-\frac{1}{2})
\end{equation}
in real space, where $\eta_{n}^{\dag}$ and $\eta_{n}$ are fermionic creation and annihilation operators, and $\Lambda_{n}$ is the excitation energy for $n^{\text{th}}$ quasiparticle mode.

The ground state is $|GS\rangle = \bigotimes_{n}|0_{n}\rangle$ in real space, where $|0_{n}\rangle (n=1,\cdots,N)$ denotes the vacuum state in the quasiparticle mode $\Lambda_{n}$, i.e., $\eta_{n}|0_{n}\rangle=0$. The ground-state energy is given by
\begin{equation}\label{ground.energy.disordered}
  E_{0} = -\sum_{n=1}^{N}\frac{1}{2}\Lambda_{n}.
\end{equation}

We study the quantum quench from $H_{0} = H(h_{0})$ to $\tilde{H} = H(h_{1})$, where the initial state $|\psi_{0}\rangle = |GS\rangle$ is taken as the ground state of the pre-quench Hamiltonian. Therefore, the time-evolved state at arbitrary time after quench is given by
\begin{equation}\label{time-evolved.state.disordered}
  |\psi(t)\rangle = e^{-i\tilde{H}t}|\psi_{0}\rangle.
\end{equation}
Considering the relation between the ground states of the pre- and post-quench\cite{Zhong201184,Cao2020102}, we have
\begin{equation}\label{superposition.disorder}
  |\psi_{0}\rangle = \frac{1}{\mathcal{N}}\exp{(\frac{1}{2}\sum_{mn}\tilde{\eta}_{m}^{\dag}
  G_{mn}\eta_{n}^{\dag})}|\tilde{\psi}_{0}\rangle.
\end{equation}
where $|\tilde{\psi}_{0}\rangle=|\tilde{GS}\rangle$ is the ground state of the post-quench Hamiltonian. Therefore, we obtain the Loschmidt amplitude and decompose $\mathcal{G}(t)=\langle\psi_{0}|\psi(t)\rangle=e^{-i\tilde{E}_{0N}t}\prod_{m=1}^{N-1}\mathcal{G}_{m}(t)$ for every quasiparticle mode $\Lambda_{m}$ with
\begin{equation}\label{Loschmidt.amplitude.disordered}
  \mathcal{G}_{m}(t)= e^{-i\tilde{E}_{0m}t}\prod_{n>m}\frac{1}{\mathcal{N}_{mn}^{2}}[1+e^{-i(\tilde{\Lambda}_{m}
  +\tilde{\Lambda}_{n})t}|G_{mn}|^{2}]
\end{equation}
in real space, where $\mathcal{N}_{mn}^{2} = 1+|G_{mn}|^{2}$ is the normalization coefficient (\textbf{see Appendix.~B.~2}). The associated Fisher zeros of the Loschmidt amplitude can be calculated by $\mathcal{G}(z)=0$, that is
\begin{equation}\label{Fisher.zeros.disordered}
  z_{j} = \frac{1}{\tilde{\Lambda}_{m}+\tilde{\Lambda}_{n}}[\ln{|G_{mn}|^{2}}
  +i(2j+1)\pi], j\in\mathbb{N}.
\end{equation}
According to Eq.~(\ref{Fisher.zeros.disordered}), we obtain the condition for the occurrence of the DQPT and the critical times as
\begin{equation}\label{critical.t.disorder}
  |G_{mn}|=1 \quad \text{and} \quad t_{c}(j) = \frac{(2j+1)\pi}{\tilde{\Lambda}_{m}+\tilde{\Lambda}_{n}}.
\end{equation}

Similar to Eq.~(\ref{polar.loschmidt.amplitude.two}), in polar coordinate, the factor of Loschmidt amplitude is given by
\begin{equation}\label{polar.loschmidt.amplitude.disorder}
  \begin{split}
    \mathcal{G}_{m}(t) & = \texttt{Re}[\mathcal{G}_{m}(t)]+i\texttt{Im}[\mathcal{G}_{m}(t)] \\
      & = r_{m}(t)e^{i\phi_{m}(t)}= r_{m}(t)e^{i[\phi_{m}^{dyn}(t)+\phi_{m}^{G}(t)]},
  \end{split}
\end{equation}
with the modulus $r_{m}(t) = \sqrt{\texttt{Re}[\mathcal{G}_{m}(t)]^{2}+
  \texttt{Im}[\mathcal{G}_{m}(t)]^{2}}$ and the argument $\phi_{m}(t) = \arg{[\mathcal{G}_{m}(t)]}$. The associated
dynamical phase $\phi_{m}^{dyn}(t)$ is
\begin{equation}\label{dynamical.phase.disorder.m}
    \phi_{m}^{dyn}(t) = -\int_{0}^{t}ds\langle\psi(s)|\tilde{H}_{m}|\psi(s)\rangle = (\frac{1}{2}-p_{m})\tilde{\Lambda}_{m}t
\end{equation}
with
\begin{equation}\label{dynamical.phase.disorder.p.m}
  p_{m} = \frac{\sum_{n>m}|G_{mn}|^{2}}{\prod_{n>m}(1+|G_{mn}|^{2})}.
\end{equation}
Therefore, the PGP in the disordered QIC can be calculated by
\begin{equation}\label{Pancharatnam.geometic.phase.m}
  \phi_{m}^{G}(t) = \phi_{m}(t) - \phi_{m}^{dyn}(t).
\end{equation}

In the following, we will show two typical examples to illustrate the PGP $\phi_{m}^{G}(t)$ in the disordered QIC with weak disorder, so that there is only one extra group of DQPTs induced by the weak disorder in the system \cite{Cao2020102}.

\subsection{Numerical Results}

\begin{figure}
  \centering
  \includegraphics[width=1.0\linewidth]{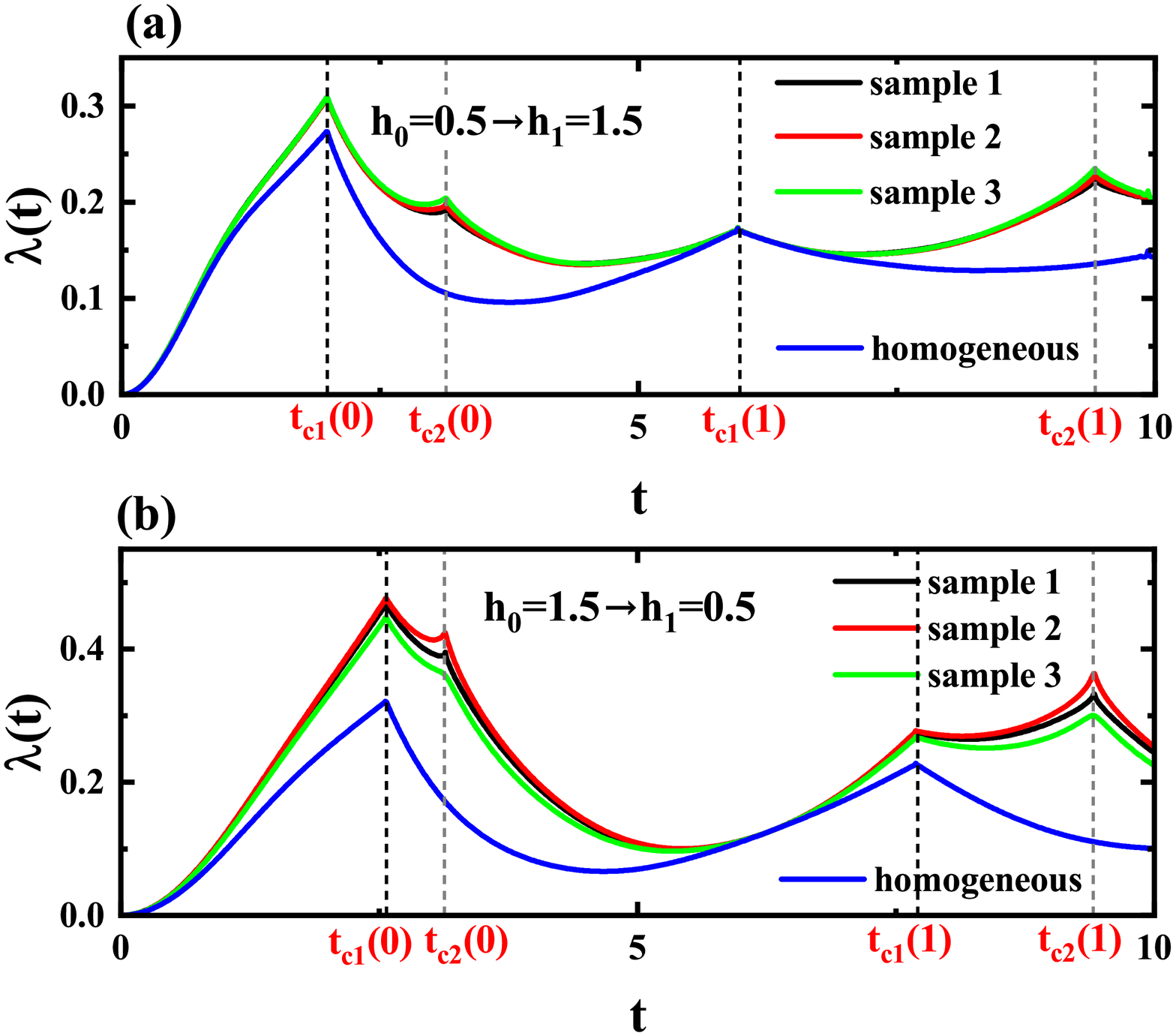}\\
  \caption{Rate functions in the disordered QIC with the strength of disorder $w=0.001$. The quench path in (a) is from $h_{0}=0.5$ to $h_{1}=1.5$, and in (b) is from $h_{0}=1.5$ to $h_{1}=0.5$. For each case, we give results of three disorder samples. It can be seen that different weakly disordered samples only influence the values of rate functions, but do not change the critical times. As comparisons, we also display the rate function of the homogeneous QIC ($w=0$). It is clear that new critical times $t_{c2}(n),n=0,1,\cdots$ of DQPTs emerge in the presence of the weak disorder. The system size is $N=1000$.}\label{ratefunction.disorder}
\end{figure}

\begin{figure}
  \centering
  \includegraphics[width=0.98\linewidth]{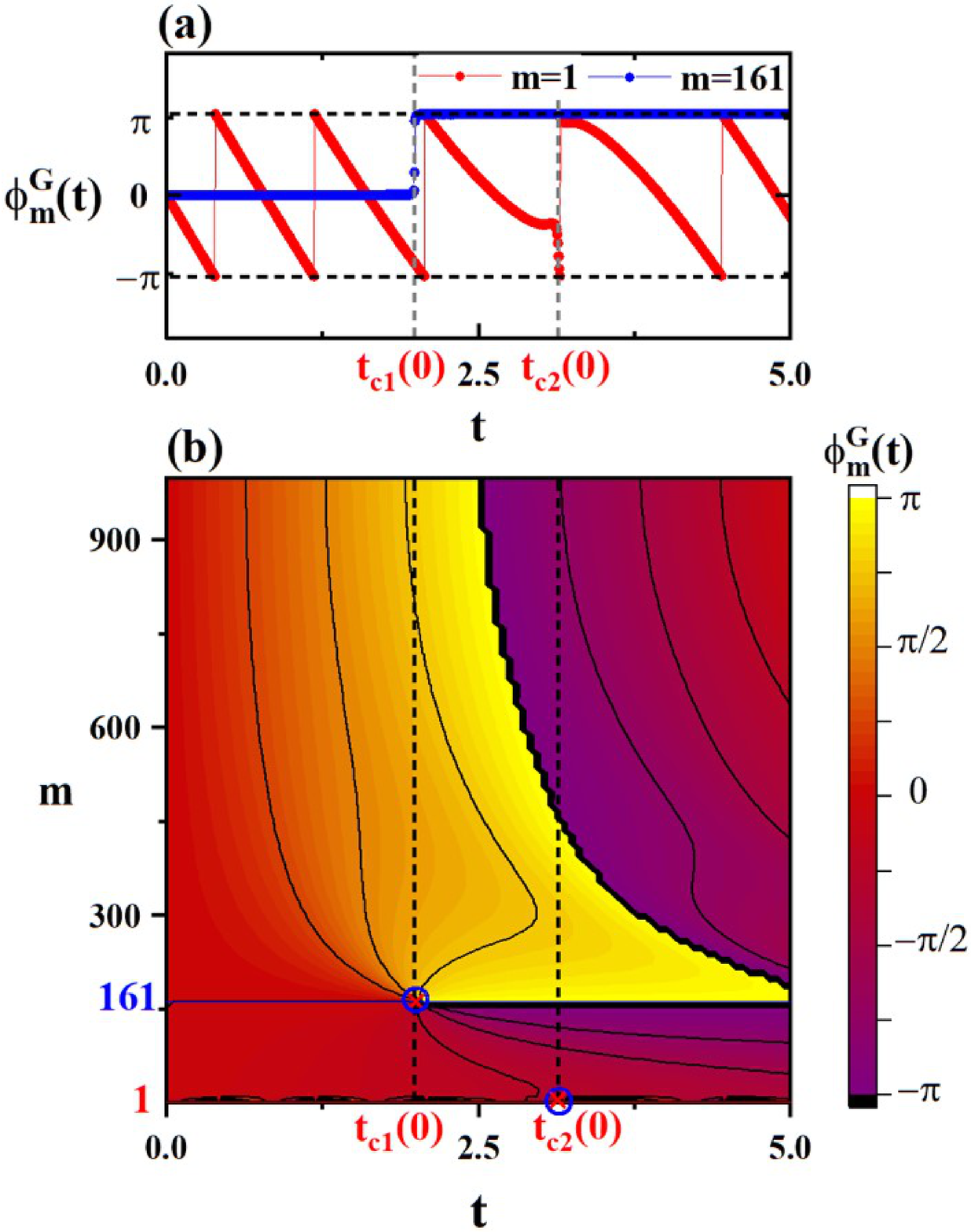}\\
  \caption{(a) The factor components of the PGP $\phi_{m}^{G}(t)$ for $m=1$ (red scatter line) and $m=161$ (blue scatter line) in the disordered QIC with $w=0.001$. The quench path is from $h_{0}=0.5$ to $h_{1}=1.5$. It can be seen that $\phi_{m=1}^{G}(t)$ and $\phi_{m=161}^{G}(t)$ show non-analytic singularity at the critical time $t_{c2}(0)\approx3.14$ and $t_{c1}(0)\approx1.99$, respectively. (b) The contour plot of the PGP $\phi_{m}^{G}(t)$ in the $(m,t)$ plane. There are two dynamical vortices, circled in blue, corresponding to the critical times $t_{c1}(0)\approx1.99$ and $t_{c2}(0)\approx3.14$. The red ``$\times$'' denotes the critical momentums and critical times obtained according to Eq.~(\ref{critical.t.disorder}).}\label{disorder.pancharatam.1}
\end{figure}

\begin{figure}
  \centering
  \includegraphics[width=1.0\linewidth]{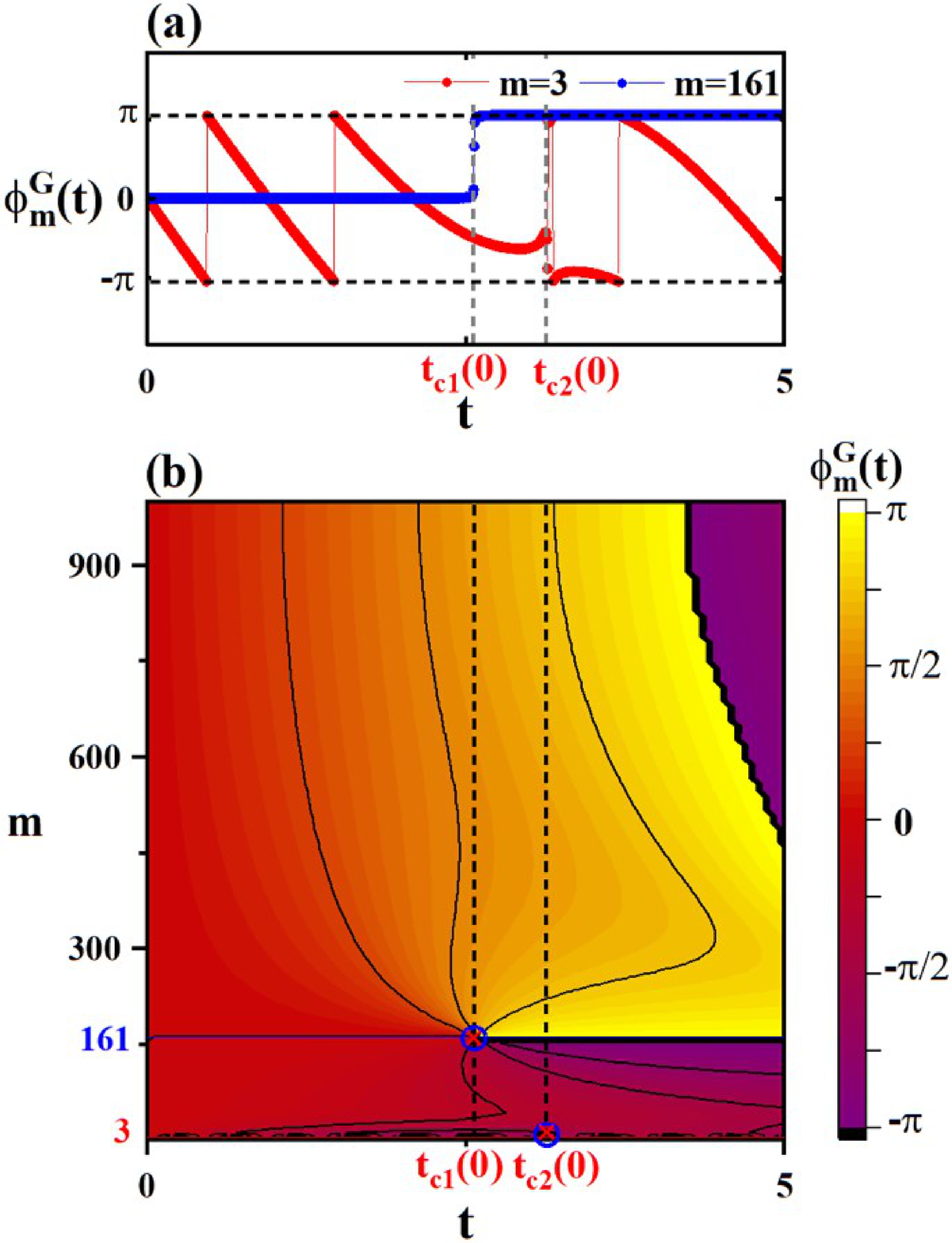}\\
  \caption{(a) The factor components of the PGP $\phi_{m}^{G}(t)$ for $m=3$ (red scatter line) and $m=161$ (blue scatter line) in the disordered QIC with $w=0.001$. The quench path is from $h_{0}=1.5$ to $h_{1}=0.5$. It can be seen that $\phi_{m=3}^{G}(t)$ and $\phi_{m=161}^{G}(t)$ show non-analytic singularity at the critical time $t_{c2}(0)\approx3.13$ and $t_{c1}(0)\approx2.57$, respectively. (b) The contour plot of the PGP $\phi_{m}^{G}(t)$ in the $(m,t)$ plane. There are two dynamical vortices, circled in blue, corresponding to the critical times $t_{c1}(0)\approx2.57$ and $t_{c2}(0)\approx3.13$. The red ``$\times$'' denotes the critical momentums and critical times obtained according to Eq.~(\ref{critical.t.disorder}). }\label{disorder.pancharatam.2}
\end{figure}

To illustrate the effect of the weak disorder on the DQPT, we show the rate functions for the weakly disordered QIC ($w=0.001$) and the homogeneous QIC ($w=0$) in Fig.~\ref{ratefunction.disorder}. The quench path in Fig.~\ref{ratefunction.disorder}~(a) is from $h_{0}=0.5$ to $h_{1}=1.5$.  It can be seen that the homogeneous QIC has the critical times $t_{c1}(n)=(2n+1)t_{c1}(0),t_{c1}(0)\approx1.99$ [see the blue line in Fig.~\ref{ratefunction.disorder}~(a)]. However, the system has one more group of critical times in the presence of weak disorder, where the new critical times induced by the disorder are given by $t_{c2}(n)=(2n+1)t_{c2}(0),t_{c2}(0)\approx3.14$.   Similar behaviors are also observed in Fig.~\ref{ratefunction.disorder}~(b), where the quench path is from $h_{0}=1.5$ to $h_{1}=0.5$. The homogeneous QIC only has one group of critical times $t_{c1}(n)=(2n+1)t_{c1}(0),t_{c1}(0)\approx2.57$, and the new extra critical times emerge in the disordered QICs, which are given by $t_{c2}(n)=(2n+1)t_{c2}(0),t_{c2}(0)\approx3.13$. Note that we display the results of three disordered samples for each quench case. It is found that the different samples only influence the values of the rate functions, but do not change the critical times of the DQPT. The critical times are generally determined by the disorder strength $w$, which has been tested for several parameters $w$. Therefore, in our work, we do not need to average over large amounts of disordered configurations, which greatly reduces our workload.

Unlike the case in the periodic QIC, the PGP can not be decomposed into every momentum $k$ due to the lack of lattice translation invariance. However, we can decompose the PGP into quasiparticle mode $\Lambda_{m}$ in real space according to Eqs.~(\ref{polar.loschmidt.amplitude.disorder}), (\ref{dynamical.phase.disorder.m}) and (\ref{Pancharatnam.geometic.phase.m}). In Fig.~\ref{disorder.pancharatam.1}, we show the contour plot of the PGP $\phi_{m}^{G}(t)$ for the quench from $h_{0}=0.5$ to $h_{1}=1.5$ in the $(m,t)$ plane, in analogy with $(k,t)$ plane in the period-two QIC. We mark the two dynamical vortices by blue circles, which are consistent with the critical times $t_{c1}(0)$ and $t_{c2}(0)$ calculated according to Eq.~(\ref{critical.t.disorder}) [see Fig.~\ref{disorder.pancharatam.1}~(b)]. Typically, the dynamical vortices are related to the non-analytic contribution to the PGP from one specific component $\phi_{m}^{G}(t)$. To find those singular components, we analyze the quasiparticle modes nearby the dynamical vortices, and find that the non-analytic point of $\phi_{m=1}^{G}(t)$ corresponds to the critical time $t_{c2}(0)$, and that of $\phi_{m=161}^{G}(t)$ to the critical time $t_{c1}(0)$ [see Fig.~\ref{disorder.pancharatam.1}~(a)].

Similarly, we study the PGP $\phi_{m}^{G}(t)$ for the quench from $h_{0}=1.5$ to $h_{1}=0.5$ [see Fig.~\ref{disorder.pancharatam.2}]. There exists two dynamical vortices at the critical times $t_{c1}(0)$ and $t_{c2}(0)$ [see Fig.~\ref{disorder.pancharatam.2}~(b)]. According to Fig.~\ref{disorder.pancharatam.2}~(a), the critical times $t_{c1}(0)$ and $t_{c2}(0)$ are induced by the non-analytic points of $\phi_{m=161}^{G}(t)$ and $\phi_{m=3}^{G}(t)$, respectively.

To summarize this section, we reformulate the PGP in real space, which allows us to study the PGP in the disordered QIC where the momentum is not a good quantum number. We observe the DQPT independently not only from the rate function, but also from the dynamical vortices of PGP in the $(m,t)$ plane. The consistency of both methods confirms the validity of our approach. It is found that the disorder induces new DQPTs in addition to those from the homogeneous QIC. Unlike the homogeneous QIC, the winding number $\nu_{D}(t)$ is not defined in the disordered QIC due to the broken of translational symmetry. We thus infer that the extra DQPTs induced by disorder are not related to the topological quantization of winding number. Recall that in the periodic QIC, the non-analytic singularities of the PGP occur when the modulus of the Loschmidt amplitude equals zero. Likewise, this is also the case in the disordered QIC.

\section{Conclusion}

In this paper, we investigate the Pancharatnam geometric phase (PGP) in the periodic and disordered QICs after a sudden quench. In the period-two QIC, we find that the winding numbers $\nu_{D}(t)$ are not quantized, and thus not topological. By comparing the results of periodic QIC with that in the homogeneous system, we clarify that the non-integer-quantized winding numbers result from the periodic modulation which can dramatically change the behavior of PGP at the boundary of the Brillouin zone. Nevertheless, the PGP still manifests non-analytic singularities at the critical times of the DQPT. This clarifies that the standard definition of the winding number can no longer serve as a topological quantum number in the period-two QIC. Furthermore, we give the general expression to calculate the PGP in real space, which allows us to investigate the PGP in the disordered QIC. Although the disorder breaks the translational invariance, we can calculate the PGP by collecting the contribution from every quasiparticle mode $\Lambda_{m}$ in real space. It is found that all the critical times, including the one induced by weak disorder, of the DQPTs in the disordered QIC have a one-to-one correspondence with the non-analytic points of the PGP. From our results, the DQPT and the non-analytic behavior of the PGP are closely related in all three cases: the homogeneous, periodic, and disordered systems, regardless of whether the winding number is quantized (topological).

Finally, we emphasize that the one-to-one correspondence between the non-analytic singularity of the PGP and the DQPT is because they both occur when the modulus $r(t)$ of the Loschmidt amplitude $\mathcal{G}(t) = r(t)e^{[i(\phi^{G}(t)+\phi^{dyn}(t)]}$ vanishes. Our work reveals the essential connection between the DQPT and the non-analytic behavior of the geometric phase, which is of great help to understand the general properties of the quantum system in the short-term dynamical process. Meanwhile, we also recognize the limitations of using the winding number as dynamical topological order parameters to describe DQPTs, which calls for a new dynamical order parameter to characterize the notion of phase and phase transitions out of equilibrium.

\begin{acknowledgments}
 K. Cao acknowledges the professor Peiqing Tong, professor Hao Guo and doctor Zhe Hou for extensive discussions and critical comments on the manuscript.
 The work is supported by National Key Basic Research Program of China (No.~2020YFB0204800), Key Research Projects of Zhejiang Lab (Nos. 2021PB0AC01 and 2021PB0AC02), and the National Science Foundation of China (Grant No.~12204432).
\end{acknowledgments}

\appendix
\section{Period-two QIC}
\subsection{Diagonalization of the period-two QIC }
For the period-two QIC (\ref{QIC.period}), by applying the Jordan-Wigner transformation $\sigma^{+}_{n}=c^{\dag}_{n}e^{i\pi\sum_{m<n}c^{\dag}_{m}c_{m}}$, $\sigma^{-}_{n}=e^{-i\pi\sum_{m<n}c^{\dag}_{m}c_{m}}c_{n}$, $\sigma^{z}_{n}=2c^{\dag}_{n}c_{n}-1$ with spin raising and lowering operators $\sigma^{\pm}_{n}=(\sigma^{x}_{n}\pm i\sigma^{y}_{n})/2$, we obtain a spinless Fermion model
\begin{equation}\label{spinless.fermion.two}
  H=-\frac{1}{2}\sum_{n=1}^{N}\{[J_{n}(c^{\dag}_{n}c_{n+1}+ c^{\dag}_{n}c^{\dag}_{n+1})+hc^{\dag}_{n}c_{n}]+h.c.\}.
\end{equation}
Note that $\{J_{n}\}$ is a period-two sequence, so that the Hamiltonian (\ref{spinless.fermion.two}) can be mapped in the complex lattices ($N'=N/2$)
\begin{equation}\label{double.lattice}
  \begin{split}
    H & =-\frac{1}{2}\sum_{n=1}^{N'}\{[J(a_{n}^{\dag}b_{n}+
               a_{n}^{\dag}b_{n}^{\dag})+ha_{n}^{\dag}a_{n}] \\
      & \quad\quad +[\alpha J(b_{n}^{\dag}a_{n+1}+b_{n}^{\dag}a_{n+1}^{\dag})+hb_{n}^{\dag}b_{n}]
        +h.c.\},
  \end{split}
\end{equation}
where $a_{2l-1}\mapsto c_{2l-1}$ and $b_{2l}\mapsto c_{2l} (l\in\mathbb{Z})$. By performing the Fourier transformation with $a_{n}=\frac{1}{\sqrt{N'}}\sum_{k\in BZ}e^{ikn}a_{k}$ and $b_{n}=\frac{1}{\sqrt{N'}}\sum_{k\in BZ}e^{ikn}b_{k}$, the resulting Hamiltonian takes the form $H=\sum_{k>0}\Psi_{k}^{\dag}H_{k}\Psi_{k}$ in momentum space $k>0$, where the spinor operator $\Psi_{k}^{\dag}=(a_{k}^{\dag},a_{-k},b_{k}^{\dag},b_{-k})$ and
\begin{widetext}
\begin{equation}\label{Bloch.Hamiltonia.two}
  H_{k}=\frac{J}{2}\left(\begin{array}{cccc}
                -2h/J & 0 & -(1+\alpha e^{-ik}) & -(1-\alpha e^{-ik}) \\
                0 & 2h/J & (1-\alpha e^{-ik}) & (1+\alpha e^{-ik}) \\
                -(1+\alpha e^{ik}) & (1-\alpha e^{ik}) & -2h/J & 0 \\
                -(1-\alpha e^{ik}) & (1+\alpha e^{ik}) & 0 & 2h/J \\
                \end{array}\right).
\end{equation}
\end{widetext}
The factor $H_{k}$ is obviously a Hermitian matrix, which can be diagonalized to the form $H_{k}=Z\Lambda Z^{\dag}$ with diagonal matrix $\Lambda_{k}=\texttt{diag}(\Lambda_{k1},-\Lambda_{k1},\Lambda_{k2},-\Lambda_{k2})$. By defining the canonical transformation
\begin{equation}\label{canonical.transformation.two}
  \Psi_{k}^{\dag}=(\eta_{k1}^{\dag},-\eta_{k1},\eta_{k2}^{\dag},-\eta_{k2}) = (a_{k}^{\dag},a_{-k},b_{k}^{\dag},b_{-k})Z,
\end{equation}
we obtain the Hamiltonian in diagonal form
\begin{equation}\label{fermions.two.appendix}
  H = \sum_{k>0}\Psi_{k}^{\dag}\Lambda_{k}\Psi_{k}.
\end{equation}

Furthermore, the canonical transformation (\ref{canonical.transformation.two}) can be expressed as
\begin{equation}\label{canonical.two}
  \left(
    \begin{array}{c}
      \Gamma_{k} \\
      \Gamma^{\dag T}_{k} \\
    \end{array}
  \right)=\left(
            \begin{array}{cc}
              U(k) & V(k) \\
              V^{*}(k) & U^{*}(k) \\
            \end{array}
          \right)\left(
                   \begin{array}{c}
                     \Phi_{k} \\
                     \Phi^{\dag T}_{k} \\
                   \end{array}
                 \right)=M\left(
                   \begin{array}{c}
                     \Phi_{k} \\
                     \Phi^{\dag T}_{k} \\
                   \end{array}
                 \right),
\end{equation}
where $\Gamma_{k}=(\eta_{k1},\eta_{-k1},\eta_{k2},\eta_{-k2})^{T}$ and $\Phi_{k}=(a_{k}, a_{-k}, b_{k}, b_{-k})^{T}$.

\subsection{Loschmidt amplitude in period-two QIC}
We study the quantum quench from $H_{0}=H(h_{0})$ to $\tilde{H}=H(h_{1})$.
According to Eq.~(\ref{canonical.two}), the canonical transformation between the quasiparticle operators of pre- and post-quench Hamiltonian is given by
\begin{equation}\label{quench.canonical.two}
  \begin{split}
    \left(
    \begin{array}{c}
      \Gamma_{k} \\
      \Gamma^{\dag T}_{k} \\
    \end{array}
  \right) & =M\tilde{M}^{-1}\left(
    \begin{array}{c}
      \tilde{\Gamma}_{k} \\
      \tilde{\Gamma}^{\dag T}_{k} \\
    \end{array}
  \right) \\
      & =\left(
           \begin{array}{cc}
             U\tilde{U}^{\dag}+V\tilde{V}^{\dag} & U\tilde{V}^{T}+V\tilde{U}^{T} \\
             U^{*}\tilde{V}^{\dag}+V^{*}\tilde{U}^{\dag} & U^{*}\tilde{U}^{T}+V^{*}\tilde{V}^{T} \\
           \end{array}
         \right)\left(
    \begin{array}{c}
      \tilde{\Gamma}_{k} \\
      \tilde{\Gamma}^{\dag T}_{k} \\
    \end{array}
  \right).
  \end{split}
\end{equation}
By considering the quasiparticle ground states satisfying $\eta_{k\mu}|\psi_{0k}\rangle=0$ and $\tilde{\eta}_{k\mu}|\tilde{\psi}_{0k}\rangle=0$, we can express the ground state $|\psi_{0k}\rangle$ of the pre-quench Hamiltonian as a superposition of the ground state $|\tilde{\psi}_{0k}\rangle$ for the post-quench Hamiltonian
\begin{equation}\label{quench.state}
  \begin{split}
    |\psi_{0k}\rangle & = \frac{1}{\mathcal{N}}\exp{[\frac{1}{2}\tilde{\Gamma}^{\dag}
  _{k}G\tilde{\Gamma}^{\dag T}_{k}]}|\tilde{\psi}_{0k}\rangle \\
      & =\frac{1}{\mathcal{N}}\prod_{\mu,\nu=1}^{2}(1+G_{k\mu,-k\nu}
      \eta_{k\mu}^{\dag}\eta_{-k\nu}^{\dag})|\tilde{\psi}_{0k}\rangle.
  \end{split}
\end{equation}
where $G=-(U\tilde{U}^{\dag}+V\tilde{V}^{\dag})^{-1}(U\tilde{V}^{T}+V\tilde{U}^{T})$.
According to the Pauli's exclusion principle of fermions and momentum conservation, the matrix $G$ have nonzero elements $G_{k\mu,-k\nu}$, that is
\begin{equation}\label{G.matrix}
  G=\left(
        \begin{array}{cccc}
          0 & G_{k1,-k1} & 0 & G_{k1,-k2}  \\
          -G_{k1,-k1} & 0 & G_{k2,-k1} & 0 \\
          0 & -G_{k2,-k1} & 0 & G_{k2,-k2} \\
          -G_{k1,-k2} & 0 & -G_{k2,-k2} & 0   \\
        \end{array}
      \right).
\end{equation}
Therefore, we can obtain the Loschmidt amplitude $\mathcal{G}(t) = \prod_{k>0}\mathcal{G}_{k}(t)$ with
\begin{equation}\label{LA.k.two.appendix}
  \mathcal{G}_{k}(t)  =\frac{e^{-i\tilde{E}_{0k}t}}{\mathcal{N}^{2}}\prod_{\mu,\nu=1}^{2}
  [1+|G_{k\mu,-k\nu}|^{2}e^{i(\tilde{\Lambda}_{k\mu}+\tilde{\Lambda}_{-k\nu})t}],
\end{equation}
where $\mathcal{N}=\prod_{\mu,\nu=1}^{2}\mathcal{N}_{k\mu,-k\nu}=\prod_{\mu,\nu=1}^{2}\sqrt{1+|G_{k\mu,-k\nu}|^{2}}$ is the normalization coefficient.

\subsection{PGP in period-two QIC}
In polar coordinate, we have
\begin{equation}\label{polar.loschmidt.amplitude.two.appendix}
  \begin{split}
    \mathcal{G}_{k}(t) & = \texttt{Re}[\mathcal{G}_{k}(t)]+i\texttt{Im}[\mathcal{G}_{k}(t)] \\
      & = r_{k}(t)e^{i\phi_{k}(t)}= r_{k}(t)e^{i[\phi_{k}^{dyn}(t)+\phi_{k}^{G}(t)]},
  \end{split}
\end{equation}
where $\texttt{Re}[\mathcal{G}_{k}(t)]$ and $\texttt{Im}[\mathcal{G}_{k}(t)]$ are the real and imaginary parts of $\mathcal{G}_{k}(t)$. Therefore, the modulus $r_{k}(t)$ of $\mathcal{G}_{k}(t)$ is given by
\begin{equation}\label{modulus.Loschmidt.amplitude.two}
  r_{k}(t) = \sqrt{\texttt{Re}[\mathcal{G}_{k}(t)]^{2}+\texttt{Im}
  [\mathcal{G}_{k}(t)]^{2}}.
\end{equation}
The argument $\phi_{k}(t)$ of $\mathcal{G}_{k}(t)$ is
\begin{equation}\label{phi.k.two}
  \phi_{k}(t)=\texttt{arg}[\mathcal{G}_{k}(t)]\in(-\pi,\pi],
\end{equation}
where we follow the standard form below
\begin{equation}\label{arg.G.k.two}
  \texttt{arg}[\mathcal{G}_{k}(t)]
  =\left\{\begin{array}{cr}
            \arctan{\frac{y}{x}}, & x>0, \\
            \frac{\pi}{2}, & x=0,y>0, \\
            -\frac{\pi}{2}, & x=0,y<0, \\
            \pi+\arctan{\frac{y}{x}}, & x<0,y>0, \\
            \arctan{\frac{y}{x}}-\pi, & x<0,y<0, \\
            0, & x>0,y=0, \\
            \pi, & x<0,y=0, \\
            \text{not defined}, & x=0.
          \end{array}
  \right.
\end{equation}
with $x=\texttt{Re}[\mathcal{G}_{k}(t)]$ and $y=\texttt{Im}[\mathcal{G}_{k}(t)]$.

The dynamical phase $\phi_{k}^{dyn}(t)$ is defined as \cite{Budich201693}
\begin{widetext}
\begin{equation}\label{dynamical.phase.two}
  \begin{split}
    \phi_{k}^{dyn} & = -\int_{0}^{t}ds\langle\psi_{k}(s)|\tilde{H}_{k}|\psi_{k}(s)\rangle   = -\int_{0}^{t}ds\langle\psi_{0k}|e^{i\tilde{H}_{k}s}\tilde{H}_{k}e^{-i\tilde{H}_{k}s}|\psi_{0k}\rangle \\
      & = -\int_{0}^{t}ds\langle\psi_{0k}|\tilde{H}_{k}|\psi_{0k}\rangle  = -t\langle\psi_{0k}|\tilde{H}_{k}|\psi_{0k}\rangle  \\
      & = -t\langle\psi_{0k}|[\tilde{\Lambda}_{k1}(\tilde{\eta}_{k1}^{\dag}\tilde{\eta}_{k1}-\frac{1}{2})
                           +\tilde{\Lambda}_{-k1}(\tilde{\eta}_{-k1}^{\dag}\tilde{\eta}_{-k1}-\frac{1}{2})
                           +\tilde{\Lambda}_{k2}(\tilde{\eta}_{k2}^{\dag}\tilde{\eta}_{k2}-\frac{1}{2})
                           +\tilde{\Lambda}_{-k2}(\tilde{\eta}_{-k2}^{\dag}\tilde{\eta}_{-k2}-\frac{1}{2})]|\psi_{0k}\rangle \\
      & = -t[\tilde{\Lambda}_{k1}\langle\psi_{0k}|\tilde{\eta}_{k1}^{\dag}\tilde{\eta}_{k1}
                                                 +\tilde{\eta}_{-k1}^{\dag}\tilde{\eta}_{-k1}|\psi_{0k}\rangle
            +\tilde{\Lambda}_{k2}\langle\psi_{0k}|\tilde{\eta}_{k2}^{\dag}\tilde{\eta}_{k2}
                                                 +\tilde{\eta}_{-k2}^{\dag}\tilde{\eta}_{-k2}|\psi_{0k}\rangle]-\tilde{E}_{0k}t,
  \end{split}
\end{equation}
where $\tilde{E}_{0k}=-(\tilde{\Lambda}_{k1}+\tilde{\Lambda}_{k2})$.
From Eq.~(\ref{quench.state}), we have
\begin{equation}\label{P.k1.k1}
  \begin{split}
    p_{k1,k1} & = \langle\psi_{0k}|\tilde{\eta}_{k1}^{\dag}\tilde{\eta}_{k1}|\psi_{0k}\rangle \\
      & = \langle\tilde{\psi}_{0k}| \frac{1}{\mathcal{N}^{2}} (1+G_{k2,-k2}^{*}\tilde{\eta}_{-k2}\tilde{\eta}_{k2})
                                                              (1+G_{-k1,k2}^{*}\tilde{\eta}_{k2}\tilde{\eta}_{-k1})
                                                              (1+G_{k1,-k2}^{*}\tilde{\eta}_{-k2}\tilde{\eta}_{k1})
                                                               \\
      & \quad\quad\quad\quad\quad
                                                 \cdot(1+G_{k1,-k1}^{*}\tilde{\eta}_{-k1}\tilde{\eta}_{k1})
             \tilde{\eta}_{k1}^{\dag}\tilde{\eta}_{k1}(1+G_{k1,-k1}\tilde{\eta}_{k1}^{\dag}\tilde{\eta}_{-k1}^{\dag})                           \\
      & \quad\quad\quad\quad\quad
        \cdot(1+G_{k1,-k2}\tilde{\eta}_{k1}^{\dag}\tilde{\eta}_{-k2}^{\dag})
             (1+G_{-k1,k2}\tilde{\eta}_{-k1}^{\dag}\tilde{\eta}_{k2}^{\dag})
             (1+G_{k2,-k2}\tilde{\eta}_{k2}^{\dag}\tilde{\eta}_{-k2}^{\dag})|\tilde{\psi}_{0k}\rangle \\
      & = \langle\tilde{\psi}_{0k}| \frac{1}{\mathcal{N}_{k1,-k1}^{2}}\frac{1}{\mathcal{N}_{k1,-k2}^{2}}
                                            (1+G_{k1,-k2}^{*}\tilde{\eta}_{-k2}\tilde{\eta}_{k1})
                                            (1+G_{k1,-k1}^{*}\tilde{\eta}_{-k1}\tilde{\eta}_{k1}) \\
      & \quad\quad\quad\quad\quad\quad\quad\cdot
          \tilde{\eta}_{k1}^{\dag}\tilde{\eta}_{k1}(1+G_{k1,-k1}\tilde{\eta}_{k1}^{\dag}\tilde{\eta}_{-k1}^{\dag})
                                                   (1+G_{k1,-k2}\tilde{\eta}_{k1}^{\dag}\tilde{\eta}_{-k2}^{\dag})
                                                   |\tilde{\psi}_{0k}\rangle \\
      & = \frac{|G_{k1,-k1}|^{2}+|G_{k1,-k2}|^{2}}{(1+|G_{k1,-k1}|^{2})(1+|G_{k1,-k2}|^{2})}.
  \end{split}
\end{equation}
\end{widetext}

\begin{figure}
  \centering
  \includegraphics[width=0.95\linewidth]{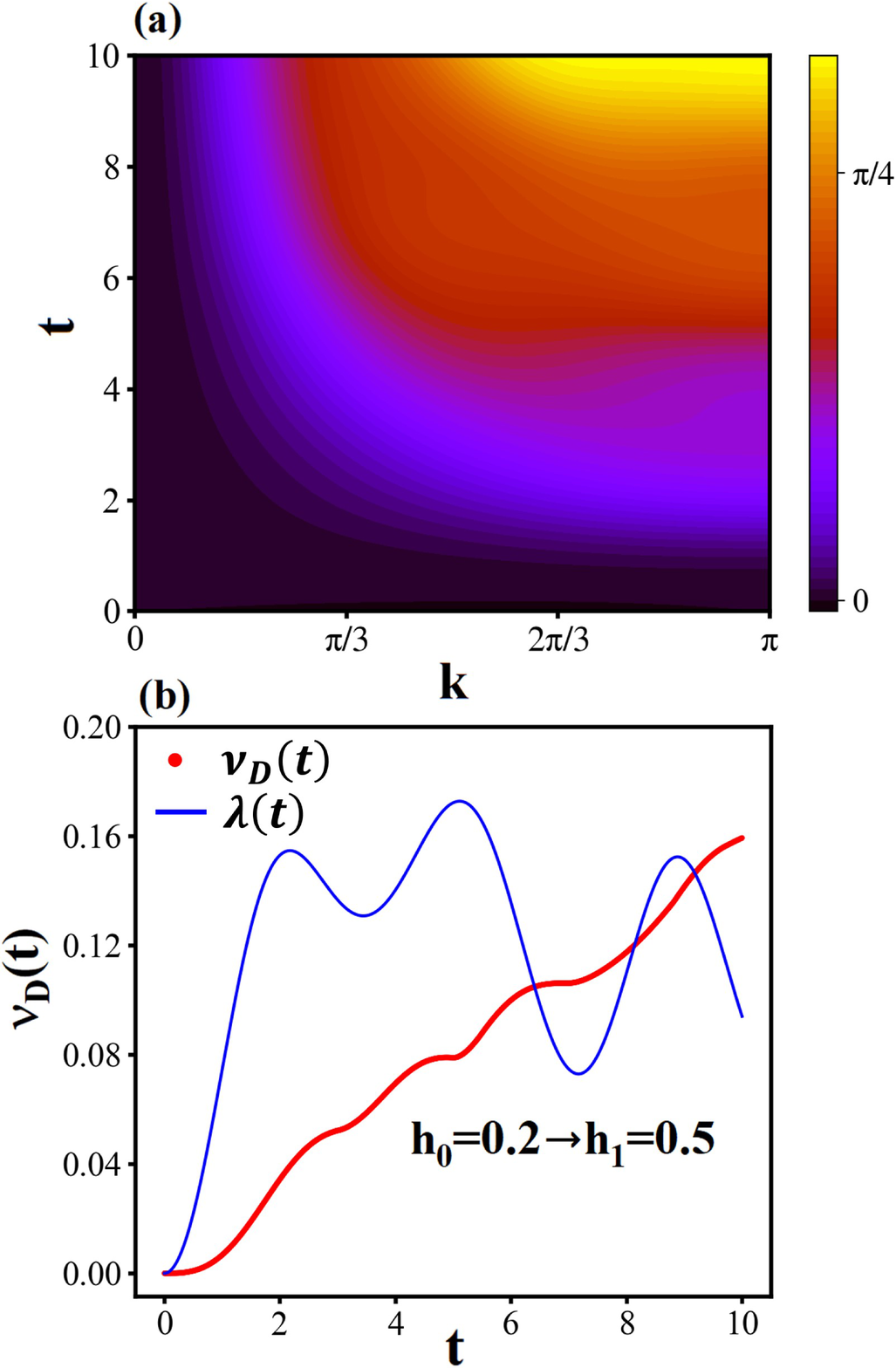}\\
  \caption{(a) Color plot of $\phi_{k}^{G}(t)$ for the quench in FM phase, without across the QPT. (b) Corresponding winding number $\nu_{D}(t)$ and rate function $\lambda(t)$.}\label{topological.same.phase}
\end{figure}

Similarly, we obtain
\begin{equation}\label{pk1k1}
  p_{k1,k1} = \frac{|G_{k1,-k1}|^{2}+|G_{k1,-k2}|^{2}}{(1+|G_{k1,-k1}|^{2})(1+|G_{k1,-k2}|^{2})},
\end{equation}
\begin{equation}\label{p-k1-k1}
  p_{-k1,-k1} = \frac{|G_{k1,-k1}|^{2}+|G_{-k1,k2}|^{2}}{(1+|G_{k1,-k1}|^{2})(1+|G_{-k1,k2}|^{2})},
\end{equation}
\begin{equation}\label{pk2k2}
  p_{k2,k2} = \frac{|G_{-k1,k2}|^{2}+|G_{k2,-k2}|^{2}}{(1+|G_{-k1,k2}|^{2})(1+|G_{k2,-k2}|^{2})},
\end{equation}
\begin{equation}\label{p-k2-k2}
  p_{-k2,-k2} = \frac{|G_{k1,-k2}|^{2}+|G_{k2,-k2}|^{2}}{(1+|G_{k1,-k2}|^{2})(1+|G_{k2,-k2}|^{2})}.
\end{equation}

By substituting Eq.(\ref{pk1k1},\ref{p-k1-k1},\ref{pk2k2},\ref{p-k2-k2}) into Eq.~(\ref{dynamical.phase.two}), we have a final formula of dynamical phase $\phi_{k}^{dyn}(t)$
\begin{equation}\label{final.dyn.phase}
  \begin{split}
    \phi_{k}^{dyn}(t) & =\{[1-\frac{2(|G_{k1,-k1}|^{2}+|G_{k1,-k2}|^{2})}{(1+|G_{k1,-k1}|^{2})(1+|G_{k1,-k2}|^{2})}]
            \tilde{\Lambda}_{k1} \\
      & + [1-\frac{2(|G_{-k1,k2}|^{2}+|G_{k2,-k2}|^{2})}{(1+|G_{-k1,k2}|^{2})(1+|G_{k2,-k2}|^{2})}]
            \tilde{\Lambda}_{k2}\} t
  \end{split}
\end{equation}
Therefore, according to Eqs.~(\ref{phi.k.two}) and (\ref{final.dyn.phase}), the PGP can be calculated by
\begin{equation}\label{Pancharatnam.phase.1}
  \phi_{k}^{G}(t) = \phi_{k}(t) - \phi_{k}^{dyn}(t).
\end{equation}

As a comparison, we show $\phi_{k}^{G}(t)$ and $\nu_{D}(t)$ for the quench in the FM phase within Fig.~\ref{topological.same.phase}. The quench path is from $h_{0}=0.2$ to $h_{1}=0.5$. It can be seen that $\phi_{k}^{G}(t)$ changes its value in the interval $[0,0.9]$ and does not have the non-analytic oscillation along with the time [see Fig.~\ref{topological.same.phase}]. Meanwhile, from Fig.~\ref{topological.same.phase}~(b), the winding number $\nu_{D}(t)$ and rate function $\lambda(t)$ are smooth continuous function along with time. This clarifies that if the quench does not cross the QPT, the dynamical topological phase transition and DQPT will not occur in the period-two QIC.

\section{Disordered QIC}

\subsection{Diagonalization of the disordered QIC}

The spin-$1/2$ quantum spin chains with nearest-neighbor interactions can be generally mapped into the spinless fermion as quadratic form via the Jordan-Wigner transformation
\begin{equation}\label{quardratic.form}
  H = \sum_{mn}[c_{m}^{\dag}A_{mn}c_{n}+\frac{1}{2}(c_{m}^{\dag}B_{mn}
  c_{n}^{\dag}+H.C.)],
\end{equation}
where $c_{n}$ and $c_{n}^{\dag}$ are the annihilation and creation operators of the fermion. For system size $N$, matrices $A$ and $B$ are both $N\times N$. Hermiticity of $H$ demands $A$ to be a Hermitian matrix and anti-commutation of fermion operators demands $B$ to be antisymmetric matrix. The matrices $A$ and $B$ are given by
\begin{eqnarray}
  A_{mn} &=& -h\delta_{mn}-J_{n}\delta_{m,n+1}/2-J_{m}\delta_{m+1,n}/2, \\
  B_{mn} &=& -J_{m}\delta_{m+1,n}/2+J_{n}\delta_{m,n+1}/2.
\end{eqnarray}

To write the Hamiltonian (\ref{quardratic.form}) in a diagonal form $H=\sum_{n}\Lambda_{n}(\eta_{n}^{\dag}\eta_{n}-\frac{1}{2})$, we can use the Bogoliubov transformation in real space
\begin{eqnarray}
  \eta_{m} &=& \sum_{n}(U_{mn}c_{n}+V_{mn}c_{n}^{\dag}), \\
  \eta_{m}^{\dag} &=& \sum_{mn}(U_{mn}^{*}c_{n}^{\dag}+V_{mn}^{*}c_{n}),
\end{eqnarray}
or in matrix form
\begin{equation}\label{Bogoliubov.disorder}
  \left(
    \begin{array}{c}
      \eta \\
      \eta^{\dag} \\
    \end{array}
  \right)=
  M\left(
     \begin{array}{c}
       c \\
       c^{\dag} \\
     \end{array}
   \right)=
   \left(
     \begin{array}{cc}
       U & V \\
       V^{*} & U^{*} \\
     \end{array}
   \right)\left(
            \begin{array}{c}
              c \\
              c^{\dag} \\
            \end{array}
          \right)
\end{equation}
with $\eta=(\eta_{1},\cdots,\eta_{N})^{T}$ and $c=(c_{1},\cdots,c_{N})^{T}$. The eigeneneries of $H$ can be obtained by solving the following eigenvalue equations:
\begin{eqnarray}
  \Phi(A-B)(A+B) &=& \Lambda^{2}\Phi, \\
  \Psi(A+B)(A-B) &=& \Lambda^{2}\Psi,
\end{eqnarray}
where $\Lambda = \texttt{diag}(\Lambda_{1},\cdots,\Lambda_{N})$. The matrices $U$ and $V$ are given by
\begin{eqnarray}
  U &=& \frac{1}{2}(\Phi + \Psi), \\
  V &=& \frac{1}{2}(\Phi - \Psi).
\end{eqnarray}

\subsection{Loschmidt amplitude in disordered QIC}

We study the quantum quench from $H_{0}=H(h_{0})$ to $\tilde{H}=H(h_{1})$. According to Eq.~(\ref{Bogoliubov.disorder}), we have
\begin{equation}\label{canonical.transformation.disordered}
  \begin{split}
    \left(
      \begin{array}{c}
        \eta \\
        \eta^{\dag} \\
      \end{array}
    \right)
     & =M\tilde{M}^{-1}\left(
                         \begin{array}{c}
                           \tilde{\eta} \\
                           \tilde{\eta}^{\dag} \\
                         \end{array}
                       \right)
      \\
      & =\left(
           \begin{array}{cc}
             U\tilde{U}^{\dag}+V\tilde{V}^{\dag} & U\tilde{V}^{T}+V\tilde{U}^{T} \\
             U^{*}\tilde{V}^{\dag}+V^{*}\tilde{U}^{\dag} & U^{*}\tilde{U}^{T}+V^{*}\tilde{V}^{T} \\
           \end{array}
         \right)\left(
                  \begin{array}{c}
                    \tilde{\eta} \\
                    \tilde{\eta}^{\dag} \\
                  \end{array}
                \right).
  \end{split}
\end{equation}
By considering $\eta_{n}|\psi_{0}\rangle=0$ and $\tilde{\eta}_{n}|\tilde{\psi}_{0}\rangle=0$, we obtain the relation between the ground states of the pre- and post-quench
\begin{equation}\label{superposition.disorder.appendix}
  \begin{split}
    |\psi_{0}\rangle & = \frac{1}{\mathcal{N}}\exp{(\frac{1}{2}\sum_{mn}\tilde{\eta}_{m}^{\dag}
  G_{mn}\eta_{n}^{\dag})}|\tilde{\psi}_{0}\rangle \\
      & = \frac{1}{\mathcal{N}}\prod_{m,n>m}(1+G_{mn}\tilde{\eta}_{m}^{\dag}
      \tilde{\eta}_{n}^{\dag})|\tilde{\psi}_{0}\rangle,
  \end{split}
\end{equation}
where $\mathcal{N}=\prod_{m,n>m}\mathcal{N}_{mn}=\prod_{m,n>m}\sqrt{1+|G_{mn}|^{2}}$ is the normalization coefficient, and  $G=-(U\tilde{U}^{\dag}+V\tilde{V}^{\dag})^{-1}(U\tilde{V}^{T}
+V\tilde{U}^{T})$ is an antisymmetrical matrix only determined by the Hamiltonian parameters. Notice that the method we calculate the Loschmidt amplitude in the disordered QIC is similar to that in the period-two QIC [see Eq.(\ref{canonical.two},\ref{quench.canonical.two},\ref{quench.state})].

According to Eq.~(\ref{superposition.disorder.appendix}), the Loschmidt amplitude is given by
\begin{equation}\label{Loschmidt.amplitude.disordered.appendix}
  \begin{split}
    \mathcal{G}(t) & = \langle\psi_{0}|\psi(t)\rangle = e^{-i\tilde{E}_{0N}t}\prod_{m=1}^{N-1}\mathcal{G}_{m}(t)
  \end{split}
\end{equation}
\begin{equation*}
  \quad\quad\quad = \frac{e^{-i\tilde{E}_{0}t}}{\mathcal{N}^{2}}\prod_{m,n>m}
      [1+e^{-i(\tilde{\Lambda}_{m}+\tilde{\Lambda}_{n})t}|G_{mn}|^{2}],
\end{equation*}
where
\begin{equation}\label{loschmidt.amplitude.disorder.m}
  \mathcal{G}_{m}(t) = e^{-i\tilde{E}_{0m}t}\prod_{n>m}\frac{1}{\mathcal{N}_{mn}^{2}}[1+e^{-i(\tilde{\Lambda}_{m}+\tilde{\Lambda}_{n})t}
  |G_{mn}|^{2}]
\end{equation}
is the component of the Loschmidt amplitude on quasiparticle mode $\tilde{\Lambda}_{m}$.

\subsection{PGP in disordered QIC chain}

In polar coordinate, we have
\begin{equation}\label{polar.loschmidt.amplitude.disorder.appendix}
  \begin{split}
    \mathcal{G}_{m}(t) & = \texttt{Re}[\mathcal{G}_{m}(t)]+i\texttt{Im}[\mathcal{G}_{m}(t)] \\
      & = r_{m}(t)e^{i\phi_{m}(t)}= r_{m}(t)e^{i[\phi_{m}^{dyn}(t)+\phi_{m}^{G}(t)]},
  \end{split}
\end{equation}
where $\texttt{Re}[\mathcal{G}_{m}(t)]$ and $\texttt{Im}[\mathcal{G}_{m}(t)]$ are the real and imaginary parts of $\mathcal{G}_{m}(t)$. Similar to the case in the period-two QIC, the modulus $r_{m}(t)$ and argument $\phi_{m}(t)$ can also be obtained by
\begin{equation}\label{modulus.disorder}
  r_{m}(t) = \sqrt{\texttt{Re}[\mathcal{G}_{m}(t)]^{2}+
  \texttt{Im}[\mathcal{G}_{m}(t)]^{2}}
\end{equation}
and
\begin{equation}\label{argument.disorder}
  \phi_{m}(t) = \arg{[\mathcal{G}_{m}(t)]},
\end{equation}
respectively.

From the definition, we have
\begin{widetext}
\begin{equation}\label{dynamical.phase.disorder}
  \begin{split}
    \phi_{m}^{dyn}(t) & = -\int_{0}^{t}ds\langle\psi(s)|\tilde{H}_{m}|\psi(s)\rangle  \\
      & = -t\langle\psi_{0}|\tilde{H}_{m}|\psi_{0}\rangle \\
      & = -t\langle\tilde{\psi}_{0}|\frac{1}{\mathcal{N}}\prod_{m',n'>m'}
      (1+G_{m'n'}^{*}\tilde{\eta}_{n'}\tilde{\eta}_{m'})
      \tilde{\Lambda}_{m}(\tilde{\eta}_{m}^{\dag}\tilde{\eta}_{m}
      -\frac{1}{2})\frac{1}{\mathcal{N}}\prod_{m',n'>m'}(1+G_{m'n'}
      \tilde{\eta}_{m'}^{\dag}\tilde{\eta}_{n'}^{\dag})|\tilde{\psi}_{0}\rangle \\
      & = \frac{1}{2}\tilde{\Lambda}_{m}t-\tilde{\Lambda}_{m}
      t\langle\tilde{\psi}_{0}|
      \frac{1}{\mathcal{N}^{2}}\prod_{m',n'>m'}(1+G_{m'n'}^{*}\tilde{\eta}_{n'}
      \tilde{\eta}_{m'})\tilde{\eta}_{m}^{\dag}\tilde{\eta}_{m}
      \prod_{m',n'>m'}(1+G_{m'n'}\tilde{\eta}_{m'}^{\dag}\tilde{\eta}_{n'}^{
      \dag})|\tilde{\psi}_{0}\rangle \\
      & = \frac{1}{2}\tilde{\Lambda}_{m}t- \tilde{\Lambda}_{m}
      t\langle\tilde{\psi}_{0}|\prod_{n>m}\frac{1}{\mathcal{N}_{m,n}^{2}}
      (1+G_{mn}^{*}\tilde{\eta}_{m}\tilde{\eta}_{n})\tilde{\eta}_{m}^{\dag}
      \tilde{\eta}_{m}\prod_{n>m}(1+G_{mn}\tilde{\eta}_{m}^{\dag}
      \tilde{\eta}_{n}^{\dag})|\tilde{\psi}_{0}\rangle \\
      & =\frac{1}{2}\tilde{\Lambda}_{m}t - \tilde{\Lambda}_{m}
      p_{m}t \\
      & = (\frac{1}{2}-p_{m})\tilde{\Lambda}_{m}t,
  \end{split}
\end{equation}
\end{widetext}
where
\begin{equation*}
  p_{1} = \frac{\sum_{n>1}|G_{1n}|^{2}}{\prod_{n>1}(1+|G_{1n}|^{2})},
\end{equation*}
\begin{equation*}
  p_{2} = \frac{\sum_{n>2}|G_{2n}|^{2}}{\prod_{n>2}(1+|G_{2n}|^{2})},
\end{equation*}
\begin{equation*}
  \vdots
\end{equation*}
\begin{equation*}
  p_{N-1} = \frac{|G_{N-1,N}|^{2}}{(1+|G_{N-1,N}|^{2})}.
\end{equation*}
Therefore, the PGPs of the disordered QIC can be calculated by
\begin{equation}\label{Pancharatnam.geometric.disorder}
  \begin{split}
    \phi_{m}^{G}(t) & = \phi_{m}(t) - \phi_{m}^{dyn}(t) \\
      & = \text{arg}[\mathcal{G}_{m}(t)]-(\frac{1}{2}-\frac{\sum_{n>m}
      |G_{mn}|^{2}}{\prod_{n>m}(1+|G_{mn}|^{2})})\tilde{\Lambda}_{m}t.
  \end{split}
\end{equation}

\bibliography{dtpt}

\providecommand{\noopsort}[1]{}\providecommand{\singleletter}[1]{#1}%
\begin{thebibliography}{90}%
\makeatletter
\providecommand \@ifxundefined [1]{%
 \@ifx{#1\undefined}
}%
\providecommand \@ifnum [1]{%
 \ifnum #1\expandafter \@firstoftwo
 \else \expandafter \@secondoftwo
 \fi
}%
\providecommand \@ifx [1]{%
 \ifx #1\expandafter \@firstoftwo
 \else \expandafter \@secondoftwo
 \fi
}%
\providecommand \natexlab [1]{#1}%
\providecommand \enquote  [1]{``#1''}%
\providecommand \bibnamefont  [1]{#1}%
\providecommand \bibfnamefont [1]{#1}%
\providecommand \citenamefont [1]{#1}%
\providecommand \href@noop [0]{\@secondoftwo}%
\providecommand \href [0]{\begingroup \@sanitize@url \@href}%
\providecommand \@href[1]{\@@startlink{#1}\@@href}%
\providecommand \@@href[1]{\endgroup#1\@@endlink}%
\providecommand \@sanitize@url [0]{\catcode `\\12\catcode `\$12\catcode
  `\&12\catcode `\#12\catcode `\^12\catcode `\_12\catcode `\%12\relax}%
\providecommand \@@startlink[1]{}%
\providecommand \@@endlink[0]{}%
\providecommand \url  [0]{\begingroup\@sanitize@url \@url }%
\providecommand \@url [1]{\endgroup\@href {#1}{\urlprefix }}%
\providecommand \urlprefix  [0]{URL }%
\providecommand \Eprint [0]{\href }%
\providecommand \doibase [0]{http://dx.doi.org/}%
\providecommand \selectlanguage [0]{\@gobble}%
\providecommand \bibinfo  [0]{\@secondoftwo}%
\providecommand \bibfield  [0]{\@secondoftwo}%
\providecommand \translation [1]{[#1]}%
\providecommand \BibitemOpen [0]{}%
\providecommand \bibitemStop [0]{}%
\providecommand \bibitemNoStop [0]{.\EOS\space}%
\providecommand \EOS [0]{\spacefactor3000\relax}%
\providecommand \BibitemShut  [1]{\csname bibitem#1\endcsname}%
\let\auto@bib@innerbib\@empty
\bibitem [{\citenamefont {Xiao}\ \emph {et~al.}(2010)\citenamefont {Xiao},
  \citenamefont {Chang},\ and\ \citenamefont {Niu}}]{Xiao201082}%
  \BibitemOpen
  \bibfield  {author} {\bibinfo {author} {\bibfnamefont {D.}~\bibnamefont
  {Xiao}}, \bibinfo {author} {\bibfnamefont {M.-C.}\ \bibnamefont {Chang}}, \
  and\ \bibinfo {author} {\bibfnamefont {Q.}~\bibnamefont {Niu}},\ }\href
  {\doibase 10.1103/RevModPhys.82.1959} {\bibfield  {journal} {\bibinfo
  {journal} {Rev. Mod. Phys.}\ }\textbf {\bibinfo {volume} {82}},\ \bibinfo
  {pages} {1959} (\bibinfo {year} {2010})}\BibitemShut {NoStop}%
\bibitem [{\citenamefont {Cooper}\ \emph {et~al.}(2019)\citenamefont {Cooper},
  \citenamefont {Dalibard},\ and\ \citenamefont {Spielman}}]{Cooper201991}%
  \BibitemOpen
  \bibfield  {author} {\bibinfo {author} {\bibfnamefont {N.~R.}\ \bibnamefont
  {Cooper}}, \bibinfo {author} {\bibfnamefont {J.}~\bibnamefont {Dalibard}}, \
  and\ \bibinfo {author} {\bibfnamefont {I.~B.}\ \bibnamefont {Spielman}},\
  }\href {\doibase 10.1103/RevModPhys.91.015005} {\bibfield  {journal}
  {\bibinfo  {journal} {Rev. Mod. Phys.}\ }\textbf {\bibinfo {volume} {91}},\
  \bibinfo {pages} {015005} (\bibinfo {year} {2019})}\BibitemShut {NoStop}%
\bibitem [{\citenamefont {Bergholtz}\ \emph {et~al.}(2021)\citenamefont
  {Bergholtz}, \citenamefont {Budich},\ and\ \citenamefont
  {Kunst}}]{Bergholtz202193}%
  \BibitemOpen
  \bibfield  {author} {\bibinfo {author} {\bibfnamefont {E.~J.}\ \bibnamefont
  {Bergholtz}}, \bibinfo {author} {\bibfnamefont {J.~C.}\ \bibnamefont
  {Budich}}, \ and\ \bibinfo {author} {\bibfnamefont {F.~K.}\ \bibnamefont
  {Kunst}},\ }\href {\doibase 10.1103/RevModPhys.93.015005} {\bibfield
  {journal} {\bibinfo  {journal} {Rev. Mod. Phys.}\ }\textbf {\bibinfo {volume}
  {93}},\ \bibinfo {pages} {015005} (\bibinfo {year} {2021})}\BibitemShut
  {NoStop}%
\bibitem [{\citenamefont {Berry}(1984)}]{Berry1984392}%
  \BibitemOpen
  \bibfield  {author} {\bibinfo {author} {\bibfnamefont {M.~V.}\ \bibnamefont
  {Berry}},\ }\href {\doibase 10.1098/rspa.1984.0023} {\bibfield  {journal}
  {\bibinfo  {journal} {Proc. R. Soc. Lond. A}\ }\textbf {\bibinfo {volume}
  {392}},\ \bibinfo {pages} {45} (\bibinfo {year} {1984})}\BibitemShut
  {NoStop}%
\bibitem [{\citenamefont {Simon}(1983)}]{Simon198351}%
  \BibitemOpen
  \bibfield  {author} {\bibinfo {author} {\bibfnamefont {B.}~\bibnamefont
  {Simon}},\ }\href {\doibase 10.1103/PhysRevLett.51.2167} {\bibfield
  {journal} {\bibinfo  {journal} {Phys. Rev. Lett.}\ }\textbf {\bibinfo
  {volume} {51}},\ \bibinfo {pages} {2167} (\bibinfo {year}
  {1983})}\BibitemShut {NoStop}%
\bibitem [{\citenamefont {Aharonov}\ and\ \citenamefont
  {Anandan}(1987)}]{Aharonov198758}%
  \BibitemOpen
  \bibfield  {author} {\bibinfo {author} {\bibfnamefont {Y.}~\bibnamefont
  {Aharonov}}\ and\ \bibinfo {author} {\bibfnamefont {J.}~\bibnamefont
  {Anandan}},\ }\href {\doibase 10.1103/PhysRevLett.58.1593} {\bibfield
  {journal} {\bibinfo  {journal} {Phys. Rev. Lett.}\ }\textbf {\bibinfo
  {volume} {58}},\ \bibinfo {pages} {1593} (\bibinfo {year}
  {1987})}\BibitemShut {NoStop}%
\bibitem [{\citenamefont {Samuel}\ and\ \citenamefont
  {Bhandari}(1988)}]{Samuel198860}%
  \BibitemOpen
  \bibfield  {author} {\bibinfo {author} {\bibfnamefont {J.}~\bibnamefont
  {Samuel}}\ and\ \bibinfo {author} {\bibfnamefont {R.}~\bibnamefont
  {Bhandari}},\ }\href {\doibase 10.1103/PhysRevLett.60.2339} {\bibfield
  {journal} {\bibinfo  {journal} {Phys. Rev. Lett.}\ }\textbf {\bibinfo
  {volume} {60}},\ \bibinfo {pages} {2339} (\bibinfo {year}
  {1988})}\BibitemShut {NoStop}%
\bibitem [{\citenamefont {Thouless}\ \emph {et~al.}(1982)\citenamefont
  {Thouless}, \citenamefont {Kohmoto}, \citenamefont {Nightingale},\ and\
  \citenamefont {den Nijs}}]{Thouless198249}%
  \BibitemOpen
  \bibfield  {author} {\bibinfo {author} {\bibfnamefont {D.~J.}\ \bibnamefont
  {Thouless}}, \bibinfo {author} {\bibfnamefont {M.}~\bibnamefont {Kohmoto}},
  \bibinfo {author} {\bibfnamefont {M.~P.}\ \bibnamefont {Nightingale}}, \ and\
  \bibinfo {author} {\bibfnamefont {M.}~\bibnamefont {den Nijs}},\ }\href
  {\doibase 10.1103/PhysRevLett.49.405} {\bibfield  {journal} {\bibinfo
  {journal} {Phys. Rev. Lett.}\ }\textbf {\bibinfo {volume} {49}},\ \bibinfo
  {pages} {405} (\bibinfo {year} {1982})}\BibitemShut {NoStop}%
\bibitem [{\citenamefont {Carollo}\ \emph {et~al.}(2020)\citenamefont
  {Carollo}, \citenamefont {Valenti},\ and\ \citenamefont
  {Spagnolo}}]{Carollo2020838}%
  \BibitemOpen
  \bibfield  {author} {\bibinfo {author} {\bibfnamefont {A.}~\bibnamefont
  {Carollo}}, \bibinfo {author} {\bibfnamefont {D.}~\bibnamefont {Valenti}}, \
  and\ \bibinfo {author} {\bibfnamefont {B.}~\bibnamefont {Spagnolo}},\ }\href
  {\doibase https://doi.org/10.1016/j.physrep.2019.11.002} {\bibfield
  {journal} {\bibinfo  {journal} {Physics Reports}\ }\textbf {\bibinfo {volume}
  {838}},\ \bibinfo {pages} {1} (\bibinfo {year} {2020})}\BibitemShut {NoStop}%
\bibitem [{\citenamefont {Zhang}\ and\ \citenamefont
  {Song}(2015)}]{Zhang2015115}%
  \BibitemOpen
  \bibfield  {author} {\bibinfo {author} {\bibfnamefont {G.}~\bibnamefont
  {Zhang}}\ and\ \bibinfo {author} {\bibfnamefont {Z.}~\bibnamefont {Song}},\
  }\href {\doibase 10.1103/PhysRevLett.115.177204} {\bibfield  {journal}
  {\bibinfo  {journal} {Phys. Rev. Lett.}\ }\textbf {\bibinfo {volume} {115}},\
  \bibinfo {pages} {177204} (\bibinfo {year} {2015})}\BibitemShut {NoStop}%
\bibitem [{\citenamefont {Budich}\ and\ \citenamefont
  {Heyl}(2016)}]{Budich201693}%
  \BibitemOpen
  \bibfield  {author} {\bibinfo {author} {\bibfnamefont {J.~C.}\ \bibnamefont
  {Budich}}\ and\ \bibinfo {author} {\bibfnamefont {M.}~\bibnamefont {Heyl}},\
  }\href {\doibase 10.1103/PhysRevB.93.085416} {\bibfield  {journal} {\bibinfo
  {journal} {Phys. Rev. B}\ }\textbf {\bibinfo {volume} {93}},\ \bibinfo
  {pages} {085416} (\bibinfo {year} {2016})}\BibitemShut {NoStop}%
\bibitem [{\citenamefont {Sharma}\ \emph {et~al.}(2016)\citenamefont {Sharma},
  \citenamefont {Divakaran}, \citenamefont {Polkovnikov},\ and\ \citenamefont
  {Dutta}}]{Sharma201693}%
  \BibitemOpen
  \bibfield  {author} {\bibinfo {author} {\bibfnamefont {S.}~\bibnamefont
  {Sharma}}, \bibinfo {author} {\bibfnamefont {U.}~\bibnamefont {Divakaran}},
  \bibinfo {author} {\bibfnamefont {A.}~\bibnamefont {Polkovnikov}}, \ and\
  \bibinfo {author} {\bibfnamefont {A.}~\bibnamefont {Dutta}},\ }\href
  {\doibase 10.1103/PhysRevB.93.144306} {\bibfield  {journal} {\bibinfo
  {journal} {Phys. Rev. B}\ }\textbf {\bibinfo {volume} {93}},\ \bibinfo
  {pages} {144306} (\bibinfo {year} {2016})}\BibitemShut {NoStop}%
\bibitem [{\citenamefont {Dutta}\ and\ \citenamefont
  {Dutta}(2017)}]{Dutta201796}%
  \BibitemOpen
  \bibfield  {author} {\bibinfo {author} {\bibfnamefont {A.}~\bibnamefont
  {Dutta}}\ and\ \bibinfo {author} {\bibfnamefont {A.}~\bibnamefont {Dutta}},\
  }\href {\doibase 10.1103/PhysRevB.96.125113} {\bibfield  {journal} {\bibinfo
  {journal} {Phys. Rev. B}\ }\textbf {\bibinfo {volume} {96}},\ \bibinfo
  {pages} {125113} (\bibinfo {year} {2017})}\BibitemShut {NoStop}%
\bibitem [{\citenamefont {Fl\"{a}schner}\ \emph {et~al.}(2017)\citenamefont
  {Fl\"{a}schner}, \citenamefont {Vogel}, \citenamefont {Tarnowski},
  \citenamefont {Rem}, \citenamefont {L\"{u}hmann}, \citenamefont {Heyl},
  \citenamefont {Budich}, \citenamefont {Mathey}, \citenamefont {Sengstock},\
  and\ \citenamefont {Weitenberg}}]{Vogel201714}%
  \BibitemOpen
  \bibfield  {author} {\bibinfo {author} {\bibfnamefont {N.}~\bibnamefont
  {Fl\"{a}schner}}, \bibinfo {author} {\bibfnamefont {D.}~\bibnamefont
  {Vogel}}, \bibinfo {author} {\bibfnamefont {M.}~\bibnamefont {Tarnowski}},
  \bibinfo {author} {\bibfnamefont {B.~S.}\ \bibnamefont {Rem}}, \bibinfo
  {author} {\bibfnamefont {D.~S.}\ \bibnamefont {L\"{u}hmann}}, \bibinfo
  {author} {\bibfnamefont {M.}~\bibnamefont {Heyl}}, \bibinfo {author}
  {\bibfnamefont {J.~C.}\ \bibnamefont {Budich}}, \bibinfo {author}
  {\bibfnamefont {L.}~\bibnamefont {Mathey}}, \bibinfo {author} {\bibfnamefont
  {K.}~\bibnamefont {Sengstock}}, \ and\ \bibinfo {author} {\bibfnamefont
  {C.}~\bibnamefont {Weitenberg}},\ }\href {\doibase 10.1038/s41567-017-0013-8}
  {\bibfield  {journal} {\bibinfo  {journal} {Nature Physics}\ }\textbf
  {\bibinfo {volume} {14}},\ \bibinfo {pages} {265} (\bibinfo {year}
  {2017})}\BibitemShut {NoStop}%
\bibitem [{\citenamefont {Heyl}\ and\ \citenamefont
  {Budich}(2017)}]{Heyl201796}%
  \BibitemOpen
  \bibfield  {author} {\bibinfo {author} {\bibfnamefont {M.}~\bibnamefont
  {Heyl}}\ and\ \bibinfo {author} {\bibfnamefont {J.~C.}\ \bibnamefont
  {Budich}},\ }\href {\doibase 10.1103/PhysRevB.96.180304} {\bibfield
  {journal} {\bibinfo  {journal} {Phys. Rev. B}\ }\textbf {\bibinfo {volume}
  {96}},\ \bibinfo {pages} {180304} (\bibinfo {year} {2017})}\BibitemShut
  {NoStop}%
\bibitem [{\citenamefont {Bhattacharjee}\ and\ \citenamefont
  {Dutta}(2018)}]{Bhattacharjee201897}%
  \BibitemOpen
  \bibfield  {author} {\bibinfo {author} {\bibfnamefont {S.}~\bibnamefont
  {Bhattacharjee}}\ and\ \bibinfo {author} {\bibfnamefont {A.}~\bibnamefont
  {Dutta}},\ }\href {\doibase 10.1103/PhysRevB.97.134306} {\bibfield  {journal}
  {\bibinfo  {journal} {Phys. Rev. B}\ }\textbf {\bibinfo {volume} {97}},\
  \bibinfo {pages} {134306} (\bibinfo {year} {2018})}\BibitemShut {NoStop}%
\bibitem [{\citenamefont {Lang}\ \emph {et~al.}(2018)\citenamefont {Lang},
  \citenamefont {Chen}, \citenamefont {Hong},\ and\ \citenamefont
  {Fan}}]{Lang201898}%
  \BibitemOpen
  \bibfield  {author} {\bibinfo {author} {\bibfnamefont {H.}~\bibnamefont
  {Lang}}, \bibinfo {author} {\bibfnamefont {Y.}~\bibnamefont {Chen}}, \bibinfo
  {author} {\bibfnamefont {Q.}~\bibnamefont {Hong}}, \ and\ \bibinfo {author}
  {\bibfnamefont {H.}~\bibnamefont {Fan}},\ }\href {\doibase
  10.1103/PhysRevB.98.134310} {\bibfield  {journal} {\bibinfo  {journal} {Phys.
  Rev. B}\ }\textbf {\bibinfo {volume} {98}},\ \bibinfo {pages} {134310}
  (\bibinfo {year} {2018})}\BibitemShut {NoStop}%
\bibitem [{\citenamefont {Qiu}\ \emph {et~al.}(2018)\citenamefont {Qiu},
  \citenamefont {Deng}, \citenamefont {Guo},\ and\ \citenamefont
  {Yi}}]{Qiu201898}%
  \BibitemOpen
  \bibfield  {author} {\bibinfo {author} {\bibfnamefont {X.}~\bibnamefont
  {Qiu}}, \bibinfo {author} {\bibfnamefont {T.-S.}\ \bibnamefont {Deng}},
  \bibinfo {author} {\bibfnamefont {G.-C.}\ \bibnamefont {Guo}}, \ and\
  \bibinfo {author} {\bibfnamefont {W.}~\bibnamefont {Yi}},\ }\href {\doibase
  10.1103/PhysRevA.98.021601} {\bibfield  {journal} {\bibinfo  {journal} {Phys.
  Rev. A}\ }\textbf {\bibinfo {volume} {98}},\ \bibinfo {pages} {021601}
  (\bibinfo {year} {2018})}\BibitemShut {NoStop}%
\bibitem [{\citenamefont {Zhou}\ \emph {et~al.}(2018)\citenamefont {Zhou},
  \citenamefont {Wang}, \citenamefont {Wang},\ and\ \citenamefont
  {Gong}}]{Zhou201898}%
  \BibitemOpen
  \bibfield  {author} {\bibinfo {author} {\bibfnamefont {L.}~\bibnamefont
  {Zhou}}, \bibinfo {author} {\bibfnamefont {Q.-h.}\ \bibnamefont {Wang}},
  \bibinfo {author} {\bibfnamefont {H.}~\bibnamefont {Wang}}, \ and\ \bibinfo
  {author} {\bibfnamefont {J.}~\bibnamefont {Gong}},\ }\href {\doibase
  10.1103/PhysRevA.98.022129} {\bibfield  {journal} {\bibinfo  {journal} {Phys.
  Rev. A}\ }\textbf {\bibinfo {volume} {98}},\ \bibinfo {pages} {022129}
  (\bibinfo {year} {2018})}\BibitemShut {NoStop}%
\bibitem [{\citenamefont {Mendl}\ and\ \citenamefont
  {Budich}(2019)}]{Mendl2019100}%
  \BibitemOpen
  \bibfield  {author} {\bibinfo {author} {\bibfnamefont {C.~B.}\ \bibnamefont
  {Mendl}}\ and\ \bibinfo {author} {\bibfnamefont {J.~C.}\ \bibnamefont
  {Budich}},\ }\href {\doibase 10.1103/PhysRevB.100.224307} {\bibfield
  {journal} {\bibinfo  {journal} {Phys. Rev. B}\ }\textbf {\bibinfo {volume}
  {100}},\ \bibinfo {pages} {224307} (\bibinfo {year} {2019})}\BibitemShut
  {NoStop}%
\bibitem [{\citenamefont {Qiu}\ \emph {et~al.}(2019)\citenamefont {Qiu},
  \citenamefont {Deng}, \citenamefont {Hu}, \citenamefont {Xue},\ and\
  \citenamefont {Yi}}]{Qiu201920}%
  \BibitemOpen
  \bibfield  {author} {\bibinfo {author} {\bibfnamefont {X.}~\bibnamefont
  {Qiu}}, \bibinfo {author} {\bibfnamefont {T.-S.}\ \bibnamefont {Deng}},
  \bibinfo {author} {\bibfnamefont {Y.}~\bibnamefont {Hu}}, \bibinfo {author}
  {\bibfnamefont {P.}~\bibnamefont {Xue}}, \ and\ \bibinfo {author}
  {\bibfnamefont {W.}~\bibnamefont {Yi}},\ }\href {\doibase
  https://doi.org/10.1016/j.isci.2019.09.037} {\bibfield  {journal} {\bibinfo
  {journal} {iScience}\ }\textbf {\bibinfo {volume} {20}},\ \bibinfo {pages}
  {392} (\bibinfo {year} {2019})}\BibitemShut {NoStop}%
\bibitem [{\citenamefont {Wang}\ \emph {et~al.}(2019)\citenamefont {Wang},
  \citenamefont {Qiu}, \citenamefont {Xiao}, \citenamefont {Zhan},
  \citenamefont {Bian}, \citenamefont {Yi},\ and\ \citenamefont
  {Xue}}]{Wang2019122}%
  \BibitemOpen
  \bibfield  {author} {\bibinfo {author} {\bibfnamefont {K.}~\bibnamefont
  {Wang}}, \bibinfo {author} {\bibfnamefont {X.}~\bibnamefont {Qiu}}, \bibinfo
  {author} {\bibfnamefont {L.}~\bibnamefont {Xiao}}, \bibinfo {author}
  {\bibfnamefont {X.}~\bibnamefont {Zhan}}, \bibinfo {author} {\bibfnamefont
  {Z.}~\bibnamefont {Bian}}, \bibinfo {author} {\bibfnamefont {W.}~\bibnamefont
  {Yi}}, \ and\ \bibinfo {author} {\bibfnamefont {P.}~\bibnamefont {Xue}},\
  }\href {\doibase 10.1103/PhysRevLett.122.020501} {\bibfield  {journal}
  {\bibinfo  {journal} {Phys. Rev. Lett.}\ }\textbf {\bibinfo {volume} {122}},\
  \bibinfo {pages} {020501} (\bibinfo {year} {2019})}\BibitemShut {NoStop}%
\bibitem [{\citenamefont {Yang}\ \emph {et~al.}(2019)\citenamefont {Yang},
  \citenamefont {Zhou}, \citenamefont {Ma}, \citenamefont {Kong}, \citenamefont
  {Wang}, \citenamefont {Qin}, \citenamefont {Rong}, \citenamefont {Wang},
  \citenamefont {Shi}, \citenamefont {Gong},\ and\ \citenamefont
  {Du}}]{Yang2019100}%
  \BibitemOpen
  \bibfield  {author} {\bibinfo {author} {\bibfnamefont {K.}~\bibnamefont
  {Yang}}, \bibinfo {author} {\bibfnamefont {L.}~\bibnamefont {Zhou}}, \bibinfo
  {author} {\bibfnamefont {W.}~\bibnamefont {Ma}}, \bibinfo {author}
  {\bibfnamefont {X.}~\bibnamefont {Kong}}, \bibinfo {author} {\bibfnamefont
  {P.}~\bibnamefont {Wang}}, \bibinfo {author} {\bibfnamefont {X.}~\bibnamefont
  {Qin}}, \bibinfo {author} {\bibfnamefont {X.}~\bibnamefont {Rong}}, \bibinfo
  {author} {\bibfnamefont {Y.}~\bibnamefont {Wang}}, \bibinfo {author}
  {\bibfnamefont {F.}~\bibnamefont {Shi}}, \bibinfo {author} {\bibfnamefont
  {J.}~\bibnamefont {Gong}}, \ and\ \bibinfo {author} {\bibfnamefont
  {J.}~\bibnamefont {Du}},\ }\href {\doibase 10.1103/PhysRevB.100.085308}
  {\bibfield  {journal} {\bibinfo  {journal} {Phys. Rev. B}\ }\textbf {\bibinfo
  {volume} {100}},\ \bibinfo {pages} {085308} (\bibinfo {year}
  {2019})}\BibitemShut {NoStop}%
\bibitem [{\citenamefont {Zache}\ \emph {et~al.}(2019)\citenamefont {Zache},
  \citenamefont {Mueller}, \citenamefont {Schneider}, \citenamefont
  {Jendrzejewski}, \citenamefont {Berges},\ and\ \citenamefont
  {Hauke}}]{Zache2019122}%
  \BibitemOpen
  \bibfield  {author} {\bibinfo {author} {\bibfnamefont {T.~V.}\ \bibnamefont
  {Zache}}, \bibinfo {author} {\bibfnamefont {N.}~\bibnamefont {Mueller}},
  \bibinfo {author} {\bibfnamefont {J.~T.}\ \bibnamefont {Schneider}}, \bibinfo
  {author} {\bibfnamefont {F.}~\bibnamefont {Jendrzejewski}}, \bibinfo {author}
  {\bibfnamefont {J.}~\bibnamefont {Berges}}, \ and\ \bibinfo {author}
  {\bibfnamefont {P.}~\bibnamefont {Hauke}},\ }\href {\doibase
  10.1103/PhysRevLett.122.050403} {\bibfield  {journal} {\bibinfo  {journal}
  {Phys. Rev. Lett.}\ }\textbf {\bibinfo {volume} {122}},\ \bibinfo {pages}
  {050403} (\bibinfo {year} {2019})}\BibitemShut {NoStop}%
\bibitem [{\citenamefont {Ding}(2020)}]{Ding2020102}%
  \BibitemOpen
  \bibfield  {author} {\bibinfo {author} {\bibfnamefont {C.}~\bibnamefont
  {Ding}},\ }\href {\doibase 10.1103/PhysRevB.102.060409} {\bibfield  {journal}
  {\bibinfo  {journal} {Phys. Rev. B}\ }\textbf {\bibinfo {volume} {102}},\
  \bibinfo {pages} {060409} (\bibinfo {year} {2020})}\BibitemShut {NoStop}%
\bibitem [{\citenamefont {Xu}\ \emph {et~al.}(2020)\citenamefont {Xu},
  \citenamefont {Wang}, \citenamefont {Heyl}, \citenamefont {Budich},
  \citenamefont {Pan}, \citenamefont {Chen}, \citenamefont {Jan}, \citenamefont
  {Sun}, \citenamefont {Xu}, \citenamefont {Han}, \citenamefont {Li},\ and\
  \citenamefont {Guo}}]{Xu20209}%
  \BibitemOpen
  \bibfield  {author} {\bibinfo {author} {\bibfnamefont {X.~Y.}\ \bibnamefont
  {Xu}}, \bibinfo {author} {\bibfnamefont {Q.~Q.}\ \bibnamefont {Wang}},
  \bibinfo {author} {\bibfnamefont {M.}~\bibnamefont {Heyl}}, \bibinfo {author}
  {\bibfnamefont {J.~C.}\ \bibnamefont {Budich}}, \bibinfo {author}
  {\bibfnamefont {W.~W.}\ \bibnamefont {Pan}}, \bibinfo {author} {\bibfnamefont
  {Z.}~\bibnamefont {Chen}}, \bibinfo {author} {\bibfnamefont {M.}~\bibnamefont
  {Jan}}, \bibinfo {author} {\bibfnamefont {K.}~\bibnamefont {Sun}}, \bibinfo
  {author} {\bibfnamefont {J.~S.}\ \bibnamefont {Xu}}, \bibinfo {author}
  {\bibfnamefont {Y.~J.}\ \bibnamefont {Han}}, \bibinfo {author} {\bibfnamefont
  {C.~F.}\ \bibnamefont {Li}}, \ and\ \bibinfo {author} {\bibfnamefont {G.~C.}\
  \bibnamefont {Guo}},\ }\href {\doibase 10.1038/s41377-019-0237-8} {\bibfield
  {journal} {\bibinfo  {journal} {Light-Science Applications}\ }\textbf
  {\bibinfo {volume} {9}} (\bibinfo {year} {2020}),\
  10.1038/s41377-019-0237-8}\BibitemShut {NoStop}%
\bibitem [{\citenamefont {Zamani}\ \emph {et~al.}(2020)\citenamefont {Zamani},
  \citenamefont {Jafari},\ and\ \citenamefont {Langari}}]{Zamani2020102}%
  \BibitemOpen
  \bibfield  {author} {\bibinfo {author} {\bibfnamefont {S.}~\bibnamefont
  {Zamani}}, \bibinfo {author} {\bibfnamefont {R.}~\bibnamefont {Jafari}}, \
  and\ \bibinfo {author} {\bibfnamefont {A.}~\bibnamefont {Langari}},\ }\href
  {\doibase 10.1103/PhysRevB.102.144306} {\bibfield  {journal} {\bibinfo
  {journal} {Phys. Rev. B}\ }\textbf {\bibinfo {volume} {102}},\ \bibinfo
  {pages} {144306} (\bibinfo {year} {2020})}\BibitemShut {NoStop}%
\bibitem [{\citenamefont {Jafari}\ and\ \citenamefont
  {Akbari}(2021)}]{Jafari2021103}%
  \BibitemOpen
  \bibfield  {author} {\bibinfo {author} {\bibfnamefont {R.}~\bibnamefont
  {Jafari}}\ and\ \bibinfo {author} {\bibfnamefont {A.}~\bibnamefont
  {Akbari}},\ }\href {\doibase 10.1103/PhysRevA.103.012204} {\bibfield
  {journal} {\bibinfo  {journal} {Phys. Rev. A}\ }\textbf {\bibinfo {volume}
  {103}},\ \bibinfo {pages} {012204} (\bibinfo {year} {2021})}\BibitemShut
  {NoStop}%
\bibitem [{\citenamefont {Sadrzadeh}\ \emph {et~al.}(2021)\citenamefont
  {Sadrzadeh}, \citenamefont {Jafari},\ and\ \citenamefont
  {Langari}}]{Sadrzadeh2021103}%
  \BibitemOpen
  \bibfield  {author} {\bibinfo {author} {\bibfnamefont {M.}~\bibnamefont
  {Sadrzadeh}}, \bibinfo {author} {\bibfnamefont {R.}~\bibnamefont {Jafari}}, \
  and\ \bibinfo {author} {\bibfnamefont {A.}~\bibnamefont {Langari}},\ }\href
  {\doibase 10.1103/PhysRevB.103.144305} {\bibfield  {journal} {\bibinfo
  {journal} {Phys. Rev. B}\ }\textbf {\bibinfo {volume} {103}},\ \bibinfo
  {pages} {144305} (\bibinfo {year} {2021})}\BibitemShut {NoStop}%
\bibitem [{\citenamefont {Yu}\ \emph {et~al.}(2021)\citenamefont {Yu},
  \citenamefont {Sacramento}, \citenamefont {Li},\ and\ \citenamefont
  {Lin}}]{Yu2021104}%
  \BibitemOpen
  \bibfield  {author} {\bibinfo {author} {\bibfnamefont {W.~C.}\ \bibnamefont
  {Yu}}, \bibinfo {author} {\bibfnamefont {P.~D.}\ \bibnamefont {Sacramento}},
  \bibinfo {author} {\bibfnamefont {Y.~C.}\ \bibnamefont {Li}}, \ and\ \bibinfo
  {author} {\bibfnamefont {H.-Q.}\ \bibnamefont {Lin}},\ }\href {\doibase
  10.1103/PhysRevB.104.085104} {\bibfield  {journal} {\bibinfo  {journal}
  {Phys. Rev. B}\ }\textbf {\bibinfo {volume} {104}},\ \bibinfo {pages}
  {085104} (\bibinfo {year} {2021})}\BibitemShut {NoStop}%
\bibitem [{\citenamefont {Zhou}\ and\ \citenamefont
  {Du}(2021{\natexlab{a}})}]{Zhou202133}%
  \BibitemOpen
  \bibfield  {author} {\bibinfo {author} {\bibfnamefont {L.}~\bibnamefont
  {Zhou}}\ and\ \bibinfo {author} {\bibfnamefont {Q.}~\bibnamefont {Du}},\
  }\href {\doibase 10.1088/1361-648x/ac0b60} {\bibfield  {journal} {\bibinfo
  {journal} {Journal of Physics: Condensed Matter}\ }\textbf {\bibinfo {volume}
  {33}},\ \bibinfo {pages} {345403} (\bibinfo {year}
  {2021}{\natexlab{a}})}\BibitemShut {NoStop}%
\bibitem [{\citenamefont {Zhou}\ and\ \citenamefont
  {Du}(2021{\natexlab{b}})}]{Zhou202123}%
  \BibitemOpen
  \bibfield  {author} {\bibinfo {author} {\bibfnamefont {L.}~\bibnamefont
  {Zhou}}\ and\ \bibinfo {author} {\bibfnamefont {Q.}~\bibnamefont {Du}},\
  }\href {\doibase 10.1088/1367-2630/ac0574} {\bibfield  {journal} {\bibinfo
  {journal} {New Journal of Physics}\ }\textbf {\bibinfo {volume} {23}},\
  \bibinfo {pages} {063041} (\bibinfo {year} {2021}{\natexlab{b}})}\BibitemShut
  {NoStop}%
\bibitem [{\citenamefont {Jafari}\ \emph {et~al.}(2022)\citenamefont {Jafari},
  \citenamefont {Akbari}, \citenamefont {Mishra},\ and\ \citenamefont
  {Johannesson}}]{Jafari2022105}%
  \BibitemOpen
  \bibfield  {author} {\bibinfo {author} {\bibfnamefont {R.}~\bibnamefont
  {Jafari}}, \bibinfo {author} {\bibfnamefont {A.}~\bibnamefont {Akbari}},
  \bibinfo {author} {\bibfnamefont {U.}~\bibnamefont {Mishra}}, \ and\ \bibinfo
  {author} {\bibfnamefont {H.}~\bibnamefont {Johannesson}},\ }\href {\doibase
  10.1103/PhysRevB.105.094311} {\bibfield  {journal} {\bibinfo  {journal}
  {Phys. Rev. B}\ }\textbf {\bibinfo {volume} {105}},\ \bibinfo {pages}
  {094311} (\bibinfo {year} {2022})}\BibitemShut {NoStop}%
\bibitem [{\citenamefont {Luan}\ \emph {et~al.}(2022)\citenamefont {Luan},
  \citenamefont {Zhang},\ and\ \citenamefont {Wang}}]{Luan2022604}%
  \BibitemOpen
  \bibfield  {author} {\bibinfo {author} {\bibfnamefont {L.-N.}\ \bibnamefont
  {Luan}}, \bibinfo {author} {\bibfnamefont {M.-Y.}\ \bibnamefont {Zhang}}, \
  and\ \bibinfo {author} {\bibfnamefont {L.}~\bibnamefont {Wang}},\ }\href
  {\doibase https://doi.org/10.1016/j.physa.2022.127866} {\bibfield  {journal}
  {\bibinfo  {journal} {Physica A: Statistical Mechanics and its Applications}\
  }\textbf {\bibinfo {volume} {604}},\ \bibinfo {pages} {127866} (\bibinfo
  {year} {2022})}\BibitemShut {NoStop}%
\bibitem [{\citenamefont {Naji}\ \emph {et~al.}(2022)\citenamefont {Naji},
  \citenamefont {Jafari}, \citenamefont {Jafari},\ and\ \citenamefont
  {Akbari}}]{Naji2022105}%
  \BibitemOpen
  \bibfield  {author} {\bibinfo {author} {\bibfnamefont {J.}~\bibnamefont
  {Naji}}, \bibinfo {author} {\bibfnamefont {M.}~\bibnamefont {Jafari}},
  \bibinfo {author} {\bibfnamefont {R.}~\bibnamefont {Jafari}}, \ and\ \bibinfo
  {author} {\bibfnamefont {A.}~\bibnamefont {Akbari}},\ }\href {\doibase
  10.1103/PhysRevA.105.022220} {\bibfield  {journal} {\bibinfo  {journal}
  {Phys. Rev. A}\ }\textbf {\bibinfo {volume} {105}},\ \bibinfo {pages}
  {022220} (\bibinfo {year} {2022})}\BibitemShut {NoStop}%
\bibitem [{\citenamefont {Heyl}\ \emph {et~al.}(2013)\citenamefont {Heyl},
  \citenamefont {Polkovnikov},\ and\ \citenamefont {Kehrein}}]{Heyl2013110}%
  \BibitemOpen
  \bibfield  {author} {\bibinfo {author} {\bibfnamefont {M.}~\bibnamefont
  {Heyl}}, \bibinfo {author} {\bibfnamefont {A.}~\bibnamefont {Polkovnikov}}, \
  and\ \bibinfo {author} {\bibfnamefont {S.}~\bibnamefont {Kehrein}},\ }\href
  {\doibase 10.1103/PhysRevLett.110.135704} {\bibfield  {journal} {\bibinfo
  {journal} {Phys. Rev. Lett.}\ }\textbf {\bibinfo {volume} {110}},\ \bibinfo
  {pages} {135704} (\bibinfo {year} {2013})}\BibitemShut {NoStop}%
\bibitem [{\citenamefont {Zvyagin}(2016)}]{Zvyagin201642}%
  \BibitemOpen
  \bibfield  {author} {\bibinfo {author} {\bibfnamefont {A.~A.}\ \bibnamefont
  {Zvyagin}},\ }\href {\doibase 10.1063/1.4969869} {\bibfield  {journal}
  {\bibinfo  {journal} {Low Temperature Physics}\ }\textbf {\bibinfo {volume}
  {42}},\ \bibinfo {pages} {971} (\bibinfo {year} {2016})}\BibitemShut
  {NoStop}%
\bibitem [{\citenamefont {Heyl}(2018)}]{Heyl201881}%
  \BibitemOpen
  \bibfield  {author} {\bibinfo {author} {\bibfnamefont {M.}~\bibnamefont
  {Heyl}},\ }\href {\doibase 10.1088/1361-6633/aaaf9a} {\bibfield  {journal}
  {\bibinfo  {journal} {Reports on Progress in Physics}\ }\textbf {\bibinfo
  {volume} {81}},\ \bibinfo {pages} {054001} (\bibinfo {year}
  {2018})}\BibitemShut {NoStop}%
\bibitem [{\citenamefont {Karrasch}\ and\ \citenamefont
  {Schuricht}(2013)}]{Karrasch201387}%
  \BibitemOpen
  \bibfield  {author} {\bibinfo {author} {\bibfnamefont {C.}~\bibnamefont
  {Karrasch}}\ and\ \bibinfo {author} {\bibfnamefont {D.}~\bibnamefont
  {Schuricht}},\ }\href {\doibase 10.1103/PhysRevB.87.195104} {\bibfield
  {journal} {\bibinfo  {journal} {Phys. Rev. B}\ }\textbf {\bibinfo {volume}
  {87}},\ \bibinfo {pages} {195104} (\bibinfo {year} {2013})}\BibitemShut
  {NoStop}%
\bibitem [{\citenamefont {Andraschko}\ and\ \citenamefont
  {Sirker}(2014)}]{Andraschko201489}%
  \BibitemOpen
  \bibfield  {author} {\bibinfo {author} {\bibfnamefont {F.}~\bibnamefont
  {Andraschko}}\ and\ \bibinfo {author} {\bibfnamefont {J.}~\bibnamefont
  {Sirker}},\ }\href {\doibase 10.1103/PhysRevB.89.125120} {\bibfield
  {journal} {\bibinfo  {journal} {Phys. Rev. B}\ }\textbf {\bibinfo {volume}
  {89}},\ \bibinfo {pages} {125120} (\bibinfo {year} {2014})}\BibitemShut
  {NoStop}%
\bibitem [{\citenamefont {Heyl}(2014)}]{Heyl2014113}%
  \BibitemOpen
  \bibfield  {author} {\bibinfo {author} {\bibfnamefont {M.}~\bibnamefont
  {Heyl}},\ }\href {\doibase 10.1103/PhysRevLett.113.205701} {\bibfield
  {journal} {\bibinfo  {journal} {Phys. Rev. Lett.}\ }\textbf {\bibinfo
  {volume} {113}},\ \bibinfo {pages} {205701} (\bibinfo {year}
  {2014})}\BibitemShut {NoStop}%
\bibitem [{\citenamefont {Hickey}\ \emph {et~al.}(2014)\citenamefont {Hickey},
  \citenamefont {Genway},\ and\ \citenamefont {Garrahan}}]{Hickey201489}%
  \BibitemOpen
  \bibfield  {author} {\bibinfo {author} {\bibfnamefont {J.~M.}\ \bibnamefont
  {Hickey}}, \bibinfo {author} {\bibfnamefont {S.}~\bibnamefont {Genway}}, \
  and\ \bibinfo {author} {\bibfnamefont {J.~P.}\ \bibnamefont {Garrahan}},\
  }\href {\doibase 10.1103/PhysRevB.89.054301} {\bibfield  {journal} {\bibinfo
  {journal} {Phys. Rev. B}\ }\textbf {\bibinfo {volume} {89}},\ \bibinfo
  {pages} {054301} (\bibinfo {year} {2014})}\BibitemShut {NoStop}%
\bibitem [{\citenamefont {Vajna}\ and\ \citenamefont
  {D\'ora}(2014)}]{Vajna201489}%
  \BibitemOpen
  \bibfield  {author} {\bibinfo {author} {\bibfnamefont {S.}~\bibnamefont
  {Vajna}}\ and\ \bibinfo {author} {\bibfnamefont {B.}~\bibnamefont {D\'ora}},\
  }\href {\doibase 10.1103/PhysRevB.89.161105} {\bibfield  {journal} {\bibinfo
  {journal} {Phys. Rev. B}\ }\textbf {\bibinfo {volume} {89}},\ \bibinfo
  {pages} {161105} (\bibinfo {year} {2014})}\BibitemShut {NoStop}%
\bibitem [{\citenamefont {Heyl}(2015)}]{Heyl2015115}%
  \BibitemOpen
  \bibfield  {author} {\bibinfo {author} {\bibfnamefont {M.}~\bibnamefont
  {Heyl}},\ }\href {\doibase 10.1103/PhysRevLett.115.140602} {\bibfield
  {journal} {\bibinfo  {journal} {Phys. Rev. Lett.}\ }\textbf {\bibinfo
  {volume} {115}},\ \bibinfo {pages} {140602} (\bibinfo {year}
  {2015})}\BibitemShut {NoStop}%
\bibitem [{\citenamefont {Schmitt}\ and\ \citenamefont
  {Kehrein}(2015)}]{Schmitt201592}%
  \BibitemOpen
  \bibfield  {author} {\bibinfo {author} {\bibfnamefont {M.}~\bibnamefont
  {Schmitt}}\ and\ \bibinfo {author} {\bibfnamefont {S.}~\bibnamefont
  {Kehrein}},\ }\href {\doibase 10.1103/PhysRevB.92.075114} {\bibfield
  {journal} {\bibinfo  {journal} {Phys. Rev. B}\ }\textbf {\bibinfo {volume}
  {92}},\ \bibinfo {pages} {075114} (\bibinfo {year} {2015})}\BibitemShut
  {NoStop}%
\bibitem [{\citenamefont {Vajna}\ and\ \citenamefont
  {D\'ora}(2015)}]{Vajna201591}%
  \BibitemOpen
  \bibfield  {author} {\bibinfo {author} {\bibfnamefont {S.}~\bibnamefont
  {Vajna}}\ and\ \bibinfo {author} {\bibfnamefont {B.}~\bibnamefont {D\'ora}},\
  }\href {\doibase 10.1103/PhysRevB.91.155127} {\bibfield  {journal} {\bibinfo
  {journal} {Phys. Rev. B}\ }\textbf {\bibinfo {volume} {91}},\ \bibinfo
  {pages} {155127} (\bibinfo {year} {2015})}\BibitemShut {NoStop}%
\bibitem [{\citenamefont {Huang}\ and\ \citenamefont
  {Balatsky}(2016)}]{Huang2016117}%
  \BibitemOpen
  \bibfield  {author} {\bibinfo {author} {\bibfnamefont {Z.}~\bibnamefont
  {Huang}}\ and\ \bibinfo {author} {\bibfnamefont {A.~V.}\ \bibnamefont
  {Balatsky}},\ }\href {\doibase 10.1103/PhysRevLett.117.086802} {\bibfield
  {journal} {\bibinfo  {journal} {Phys. Rev. Lett.}\ }\textbf {\bibinfo
  {volume} {117}},\ \bibinfo {pages} {086802} (\bibinfo {year}
  {2016})}\BibitemShut {NoStop}%
\bibitem [{\citenamefont {Bhattacharya}\ \emph {et~al.}(2017)\citenamefont
  {Bhattacharya}, \citenamefont {Bandyopadhyay},\ and\ \citenamefont
  {Dutta}}]{Bhattacharya201796}%
  \BibitemOpen
  \bibfield  {author} {\bibinfo {author} {\bibfnamefont {U.}~\bibnamefont
  {Bhattacharya}}, \bibinfo {author} {\bibfnamefont {S.}~\bibnamefont
  {Bandyopadhyay}}, \ and\ \bibinfo {author} {\bibfnamefont {A.}~\bibnamefont
  {Dutta}},\ }\href {\doibase 10.1103/PhysRevB.96.180303} {\bibfield  {journal}
  {\bibinfo  {journal} {Phys. Rev. B}\ }\textbf {\bibinfo {volume} {96}},\
  \bibinfo {pages} {180303} (\bibinfo {year} {2017})}\BibitemShut {NoStop}%
\bibitem [{\citenamefont {Bhattacharya}\ and\ \citenamefont
  {Dutta}(2017)}]{Bhattacharya201796014302}%
  \BibitemOpen
  \bibfield  {author} {\bibinfo {author} {\bibfnamefont {U.}~\bibnamefont
  {Bhattacharya}}\ and\ \bibinfo {author} {\bibfnamefont {A.}~\bibnamefont
  {Dutta}},\ }\href {\doibase 10.1103/PhysRevB.96.014302} {\bibfield  {journal}
  {\bibinfo  {journal} {Phys. Rev. B}\ }\textbf {\bibinfo {volume} {96}},\
  \bibinfo {pages} {014302} (\bibinfo {year} {2017})}\BibitemShut {NoStop}%
\bibitem [{\citenamefont {G\'omez-Le\'on}\ and\ \citenamefont
  {Stamp}(2017)}]{Stamp201795}%
  \BibitemOpen
  \bibfield  {author} {\bibinfo {author} {\bibfnamefont {A.}~\bibnamefont
  {G\'omez-Le\'on}}\ and\ \bibinfo {author} {\bibfnamefont {P.~C.~E.}\
  \bibnamefont {Stamp}},\ }\href {\doibase 10.1103/PhysRevB.95.054402}
  {\bibfield  {journal} {\bibinfo  {journal} {Phys. Rev. B}\ }\textbf {\bibinfo
  {volume} {95}},\ \bibinfo {pages} {054402} (\bibinfo {year}
  {2017})}\BibitemShut {NoStop}%
\bibitem [{\citenamefont {Halimeh}\ and\ \citenamefont
  {Zauner-Stauber}(2017)}]{Halimeh201796}%
  \BibitemOpen
  \bibfield  {author} {\bibinfo {author} {\bibfnamefont {J.~C.}\ \bibnamefont
  {Halimeh}}\ and\ \bibinfo {author} {\bibfnamefont {V.}~\bibnamefont
  {Zauner-Stauber}},\ }\href {\doibase 10.1103/PhysRevB.96.134427} {\bibfield
  {journal} {\bibinfo  {journal} {Phys. Rev. B}\ }\textbf {\bibinfo {volume}
  {96}},\ \bibinfo {pages} {134427} (\bibinfo {year} {2017})}\BibitemShut
  {NoStop}%
\bibitem [{\citenamefont {Homrighausen}\ \emph {et~al.}(2017)\citenamefont
  {Homrighausen}, \citenamefont {Abeling}, \citenamefont {Zauner-Stauber},\
  and\ \citenamefont {Halimeh}}]{Homrighausen201796}%
  \BibitemOpen
  \bibfield  {author} {\bibinfo {author} {\bibfnamefont {I.}~\bibnamefont
  {Homrighausen}}, \bibinfo {author} {\bibfnamefont {N.~O.}\ \bibnamefont
  {Abeling}}, \bibinfo {author} {\bibfnamefont {V.}~\bibnamefont
  {Zauner-Stauber}}, \ and\ \bibinfo {author} {\bibfnamefont {J.~C.}\
  \bibnamefont {Halimeh}},\ }\href {\doibase 10.1103/PhysRevB.96.104436}
  {\bibfield  {journal} {\bibinfo  {journal} {Phys. Rev. B}\ }\textbf {\bibinfo
  {volume} {96}},\ \bibinfo {pages} {104436} (\bibinfo {year}
  {2017})}\BibitemShut {NoStop}%
\bibitem [{\citenamefont {Weidinger}\ \emph {et~al.}(2017)\citenamefont
  {Weidinger}, \citenamefont {Heyl}, \citenamefont {Silva},\ and\ \citenamefont
  {Knap}}]{Weidinger201796}%
  \BibitemOpen
  \bibfield  {author} {\bibinfo {author} {\bibfnamefont {S.~A.}\ \bibnamefont
  {Weidinger}}, \bibinfo {author} {\bibfnamefont {M.}~\bibnamefont {Heyl}},
  \bibinfo {author} {\bibfnamefont {A.}~\bibnamefont {Silva}}, \ and\ \bibinfo
  {author} {\bibfnamefont {M.}~\bibnamefont {Knap}},\ }\href {\doibase
  10.1103/PhysRevB.96.134313} {\bibfield  {journal} {\bibinfo  {journal} {Phys.
  Rev. B}\ }\textbf {\bibinfo {volume} {96}},\ \bibinfo {pages} {134313}
  (\bibinfo {year} {2017})}\BibitemShut {NoStop}%
\bibitem [{\citenamefont {Yang}\ \emph {et~al.}(2017)\citenamefont {Yang},
  \citenamefont {Wang}, \citenamefont {Wang}, \citenamefont {Gao},\ and\
  \citenamefont {Chen}}]{Yang201796}%
  \BibitemOpen
  \bibfield  {author} {\bibinfo {author} {\bibfnamefont {C.}~\bibnamefont
  {Yang}}, \bibinfo {author} {\bibfnamefont {Y.}~\bibnamefont {Wang}}, \bibinfo
  {author} {\bibfnamefont {P.}~\bibnamefont {Wang}}, \bibinfo {author}
  {\bibfnamefont {X.}~\bibnamefont {Gao}}, \ and\ \bibinfo {author}
  {\bibfnamefont {S.}~\bibnamefont {Chen}},\ }\href {\doibase
  10.1103/PhysRevB.95.184201} {\bibfield  {journal} {\bibinfo  {journal} {Phys.
  Rev. B}\ }\textbf {\bibinfo {volume} {95}},\ \bibinfo {pages} {184201}
  (\bibinfo {year} {2017})}\BibitemShut {NoStop}%
\bibitem [{\citenamefont {Kosior}\ and\ \citenamefont
  {Sacha}(2018)}]{Kosior201897}%
  \BibitemOpen
  \bibfield  {author} {\bibinfo {author} {\bibfnamefont {A.}~\bibnamefont
  {Kosior}}\ and\ \bibinfo {author} {\bibfnamefont {K.}~\bibnamefont {Sacha}},\
  }\href {\doibase 10.1103/PhysRevA.97.053621} {\bibfield  {journal} {\bibinfo
  {journal} {Phys. Rev. A}\ }\textbf {\bibinfo {volume} {97}},\ \bibinfo
  {pages} {053621} (\bibinfo {year} {2018})}\BibitemShut {NoStop}%
\bibitem [{\citenamefont {Mera}\ \emph {et~al.}(2018)\citenamefont {Mera},
  \citenamefont {Vlachou}, \citenamefont {Paunkovi\ifmmode~\acute{c}\else
  \'{c}\fi{}}, \citenamefont {Vieira},\ and\ \citenamefont
  {Viyuela}}]{Mera201897}%
  \BibitemOpen
  \bibfield  {author} {\bibinfo {author} {\bibfnamefont {B.}~\bibnamefont
  {Mera}}, \bibinfo {author} {\bibfnamefont {C.}~\bibnamefont {Vlachou}},
  \bibinfo {author} {\bibfnamefont {N.}~\bibnamefont
  {Paunkovi\ifmmode~\acute{c}\else \'{c}\fi{}}}, \bibinfo {author}
  {\bibfnamefont {V.~R.}\ \bibnamefont {Vieira}}, \ and\ \bibinfo {author}
  {\bibfnamefont {O.}~\bibnamefont {Viyuela}},\ }\href {\doibase
  10.1103/PhysRevB.97.094110} {\bibfield  {journal} {\bibinfo  {journal} {Phys.
  Rev. B}\ }\textbf {\bibinfo {volume} {97}},\ \bibinfo {pages} {094110}
  (\bibinfo {year} {2018})}\BibitemShut {NoStop}%
\bibitem [{\citenamefont {\ifmmode \check{Z}\else
  \v{Z}\fi{}unkovi\ifmmode~\check{c}\else \v{c}\fi{}}\ \emph
  {et~al.}(2018)\citenamefont {\ifmmode \check{Z}\else
  \v{Z}\fi{}unkovi\ifmmode~\check{c}\else \v{c}\fi{}}, \citenamefont {Heyl},
  \citenamefont {Knap},\ and\ \citenamefont {Silva}}]{Bojan2018120}%
  \BibitemOpen
  \bibfield  {author} {\bibinfo {author} {\bibfnamefont {B.}~\bibnamefont
  {\ifmmode \check{Z}\else \v{Z}\fi{}unkovi\ifmmode~\check{c}\else
  \v{c}\fi{}}}, \bibinfo {author} {\bibfnamefont {M.}~\bibnamefont {Heyl}},
  \bibinfo {author} {\bibfnamefont {M.}~\bibnamefont {Knap}}, \ and\ \bibinfo
  {author} {\bibfnamefont {A.}~\bibnamefont {Silva}},\ }\href {\doibase
  10.1103/PhysRevLett.120.130601} {\bibfield  {journal} {\bibinfo  {journal}
  {Phys. Rev. Lett.}\ }\textbf {\bibinfo {volume} {120}},\ \bibinfo {pages}
  {130601} (\bibinfo {year} {2018})}\BibitemShut {NoStop}%
\bibitem [{\citenamefont {Abdi}(2019)}]{Mehdi2019100}%
  \BibitemOpen
  \bibfield  {author} {\bibinfo {author} {\bibfnamefont {M.}~\bibnamefont
  {Abdi}},\ }\href {\doibase 10.1103/PhysRevB.100.184310} {\bibfield  {journal}
  {\bibinfo  {journal} {Phys. Rev. B}\ }\textbf {\bibinfo {volume} {100}},\
  \bibinfo {pages} {184310} (\bibinfo {year} {2019})}\BibitemShut {NoStop}%
\bibitem [{\citenamefont {Huang}\ \emph {et~al.}(2019)\citenamefont {Huang},
  \citenamefont {Banerjee},\ and\ \citenamefont {Heyl}}]{Huang2019122}%
  \BibitemOpen
  \bibfield  {author} {\bibinfo {author} {\bibfnamefont {Y.-P.}\ \bibnamefont
  {Huang}}, \bibinfo {author} {\bibfnamefont {D.}~\bibnamefont {Banerjee}}, \
  and\ \bibinfo {author} {\bibfnamefont {M.}~\bibnamefont {Heyl}},\ }\href
  {\doibase 10.1103/PhysRevLett.122.250401} {\bibfield  {journal} {\bibinfo
  {journal} {Phys. Rev. Lett.}\ }\textbf {\bibinfo {volume} {122}},\ \bibinfo
  {pages} {250401} (\bibinfo {year} {2019})}\BibitemShut {NoStop}%
\bibitem [{\citenamefont {Lahiri}\ and\ \citenamefont
  {Bera}(2019)}]{Lahiri201999}%
  \BibitemOpen
  \bibfield  {author} {\bibinfo {author} {\bibfnamefont {A.}~\bibnamefont
  {Lahiri}}\ and\ \bibinfo {author} {\bibfnamefont {S.}~\bibnamefont {Bera}},\
  }\href {\doibase 10.1103/PhysRevB.99.174311} {\bibfield  {journal} {\bibinfo
  {journal} {Phys. Rev. B}\ }\textbf {\bibinfo {volume} {99}},\ \bibinfo
  {pages} {174311} (\bibinfo {year} {2019})}\BibitemShut {NoStop}%
\bibitem [{\citenamefont {Liu}\ and\ \citenamefont {Guo}(2019)}]{Liu201999}%
  \BibitemOpen
  \bibfield  {author} {\bibinfo {author} {\bibfnamefont {T.}~\bibnamefont
  {Liu}}\ and\ \bibinfo {author} {\bibfnamefont {H.}~\bibnamefont {Guo}},\
  }\href {\doibase 10.1103/PhysRevB.99.104307} {\bibfield  {journal} {\bibinfo
  {journal} {Phys. Rev. B}\ }\textbf {\bibinfo {volume} {99}},\ \bibinfo
  {pages} {104307} (\bibinfo {year} {2019})}\BibitemShut {NoStop}%
\bibitem [{\citenamefont {Cao}\ \emph {et~al.}(2020)\citenamefont {Cao},
  \citenamefont {Li}, \citenamefont {Zhong},\ and\ \citenamefont
  {Tong}}]{Cao2020102}%
  \BibitemOpen
  \bibfield  {author} {\bibinfo {author} {\bibfnamefont {K.}~\bibnamefont
  {Cao}}, \bibinfo {author} {\bibfnamefont {W.}~\bibnamefont {Li}}, \bibinfo
  {author} {\bibfnamefont {M.}~\bibnamefont {Zhong}}, \ and\ \bibinfo {author}
  {\bibfnamefont {P.}~\bibnamefont {Tong}},\ }\href {\doibase
  10.1103/PhysRevB.102.014207} {\bibfield  {journal} {\bibinfo  {journal}
  {Phys. Rev. B}\ }\textbf {\bibinfo {volume} {102}},\ \bibinfo {pages}
  {014207} (\bibinfo {year} {2020})}\BibitemShut {NoStop}%
\bibitem [{\citenamefont {Haldar}\ \emph {et~al.}(2020)\citenamefont {Haldar},
  \citenamefont {Roy}, \citenamefont {Chanda}, \citenamefont {Sen(De)},\ and\
  \citenamefont {Sen}}]{Haldar2020101}%
  \BibitemOpen
  \bibfield  {author} {\bibinfo {author} {\bibfnamefont {S.}~\bibnamefont
  {Haldar}}, \bibinfo {author} {\bibfnamefont {S.}~\bibnamefont {Roy}},
  \bibinfo {author} {\bibfnamefont {T.}~\bibnamefont {Chanda}}, \bibinfo
  {author} {\bibfnamefont {A.}~\bibnamefont {Sen(De)}}, \ and\ \bibinfo
  {author} {\bibfnamefont {U.}~\bibnamefont {Sen}},\ }\href {\doibase
  10.1103/PhysRevB.101.224304} {\bibfield  {journal} {\bibinfo  {journal}
  {Phys. Rev. B}\ }\textbf {\bibinfo {volume} {101}},\ \bibinfo {pages}
  {224304} (\bibinfo {year} {2020})}\BibitemShut {NoStop}%
\bibitem [{\citenamefont {Kyaw}\ \emph {et~al.}(2020)\citenamefont {Kyaw},
  \citenamefont {Bastidas}, \citenamefont {Tangpanitanon}, \citenamefont
  {Romero},\ and\ \citenamefont {Kwek}}]{Kyaw2020101}%
  \BibitemOpen
  \bibfield  {author} {\bibinfo {author} {\bibfnamefont {T.~H.}\ \bibnamefont
  {Kyaw}}, \bibinfo {author} {\bibfnamefont {V.~M.}\ \bibnamefont {Bastidas}},
  \bibinfo {author} {\bibfnamefont {J.}~\bibnamefont {Tangpanitanon}}, \bibinfo
  {author} {\bibfnamefont {G.}~\bibnamefont {Romero}}, \ and\ \bibinfo {author}
  {\bibfnamefont {L.-C.}\ \bibnamefont {Kwek}},\ }\href {\doibase
  10.1103/PhysRevA.101.012111} {\bibfield  {journal} {\bibinfo  {journal}
  {Phys. Rev. A}\ }\textbf {\bibinfo {volume} {101}},\ \bibinfo {pages}
  {012111} (\bibinfo {year} {2020})}\BibitemShut {NoStop}%
\bibitem [{\citenamefont {Wu}(2020)}]{Wu2020101}%
  \BibitemOpen
  \bibfield  {author} {\bibinfo {author} {\bibfnamefont {Y.}~\bibnamefont
  {Wu}},\ }\href {\doibase 10.1103/PhysRevB.101.064427} {\bibfield  {journal}
  {\bibinfo  {journal} {Phys. Rev. B}\ }\textbf {\bibinfo {volume} {101}},\
  \bibinfo {pages} {064427} (\bibinfo {year} {2020})}\BibitemShut {NoStop}%
\bibitem [{\citenamefont {Halimeh}\ \emph {et~al.}(2021)\citenamefont
  {Halimeh}, \citenamefont {Van~Damme}, \citenamefont {Guo}, \citenamefont
  {Lang},\ and\ \citenamefont {Hauke}}]{Halimeh2021104}%
  \BibitemOpen
  \bibfield  {author} {\bibinfo {author} {\bibfnamefont {J.~C.}\ \bibnamefont
  {Halimeh}}, \bibinfo {author} {\bibfnamefont {M.}~\bibnamefont {Van~Damme}},
  \bibinfo {author} {\bibfnamefont {L.}~\bibnamefont {Guo}}, \bibinfo {author}
  {\bibfnamefont {J.}~\bibnamefont {Lang}}, \ and\ \bibinfo {author}
  {\bibfnamefont {P.}~\bibnamefont {Hauke}},\ }\href {\doibase
  10.1103/PhysRevB.104.115133} {\bibfield  {journal} {\bibinfo  {journal}
  {Phys. Rev. B}\ }\textbf {\bibinfo {volume} {104}},\ \bibinfo {pages}
  {115133} (\bibinfo {year} {2021})}\BibitemShut {NoStop}%
\bibitem [{\citenamefont {Modak}\ and\ \citenamefont
  {Rakshit}(2021)}]{Modak2021103}%
  \BibitemOpen
  \bibfield  {author} {\bibinfo {author} {\bibfnamefont {R.}~\bibnamefont
  {Modak}}\ and\ \bibinfo {author} {\bibfnamefont {D.}~\bibnamefont
  {Rakshit}},\ }\href {\doibase 10.1103/PhysRevB.103.224310} {\bibfield
  {journal} {\bibinfo  {journal} {Phys. Rev. B}\ }\textbf {\bibinfo {volume}
  {103}},\ \bibinfo {pages} {224310} (\bibinfo {year} {2021})}\BibitemShut
  {NoStop}%
\bibitem [{\citenamefont {Cao}\ \emph {et~al.}(2022{\natexlab{a}})\citenamefont
  {Cao}, \citenamefont {Zhong},\ and\ \citenamefont {Tong}}]{Cao202231}%
  \BibitemOpen
  \bibfield  {author} {\bibinfo {author} {\bibfnamefont {K.}~\bibnamefont
  {Cao}}, \bibinfo {author} {\bibfnamefont {M.}~\bibnamefont {Zhong}}, \ and\
  \bibinfo {author} {\bibfnamefont {P.}~\bibnamefont {Tong}},\ }\href {\doibase
  10.1088/1674-1056/ac4a6e} {\bibfield  {journal} {\bibinfo  {journal} {Chinese
  Physics B}\ }\textbf {\bibinfo {volume} {31}},\ \bibinfo {pages} {060505}
  (\bibinfo {year} {2022}{\natexlab{a}})}\BibitemShut {NoStop}%
\bibitem [{\citenamefont {Jensen}\ \emph {et~al.}(2022)\citenamefont {Jensen},
  \citenamefont {Pedersen},\ and\ \citenamefont {Zinner}}]{Jensen2022105}%
  \BibitemOpen
  \bibfield  {author} {\bibinfo {author} {\bibfnamefont {R.~B.}\ \bibnamefont
  {Jensen}}, \bibinfo {author} {\bibfnamefont {S.~P.}\ \bibnamefont
  {Pedersen}}, \ and\ \bibinfo {author} {\bibfnamefont {N.~T.}\ \bibnamefont
  {Zinner}},\ }\href {\doibase 10.1103/PhysRevB.105.224309} {\bibfield
  {journal} {\bibinfo  {journal} {Phys. Rev. B}\ }\textbf {\bibinfo {volume}
  {105}},\ \bibinfo {pages} {224309} (\bibinfo {year} {2022})}\BibitemShut
  {NoStop}%
\bibitem [{\citenamefont {Wrze\ifmmode~\acute{s}\else \'{s}\fi{}niewski}\ \emph
  {et~al.}(2022)\citenamefont {Wrze\ifmmode~\acute{s}\else \'{s}\fi{}niewski},
  \citenamefont {Weymann}, \citenamefont {Sedlmayr},\ and\ \citenamefont
  {Doma\ifmmode~\acute{n}\else \'{n}\fi{}ski}}]{Weymann2022105}%
  \BibitemOpen
  \bibfield  {author} {\bibinfo {author} {\bibfnamefont {K.}~\bibnamefont
  {Wrze\ifmmode~\acute{s}\else \'{s}\fi{}niewski}}, \bibinfo {author}
  {\bibfnamefont {I.}~\bibnamefont {Weymann}}, \bibinfo {author} {\bibfnamefont
  {N.}~\bibnamefont {Sedlmayr}}, \ and\ \bibinfo {author} {\bibfnamefont
  {T.}~\bibnamefont {Doma\ifmmode~\acute{n}\else \'{n}\fi{}ski}},\ }\href
  {\doibase 10.1103/PhysRevB.105.094514} {\bibfield  {journal} {\bibinfo
  {journal} {Phys. Rev. B}\ }\textbf {\bibinfo {volume} {105}},\ \bibinfo
  {pages} {094514} (\bibinfo {year} {2022})}\BibitemShut {NoStop}%
\bibitem [{\citenamefont {Hou}\ \emph {et~al.}(2022)\citenamefont {Hou},
  \citenamefont {Gao}, \citenamefont {Guo},\ and\ \citenamefont
  {Chien}}]{Hou2022106}%
  \BibitemOpen
  \bibfield  {author} {\bibinfo {author} {\bibfnamefont {X.-Y.}\ \bibnamefont
  {Hou}}, \bibinfo {author} {\bibfnamefont {Q.-C.}\ \bibnamefont {Gao}},
  \bibinfo {author} {\bibfnamefont {H.}~\bibnamefont {Guo}}, \ and\ \bibinfo
  {author} {\bibfnamefont {C.-C.}\ \bibnamefont {Chien}},\ }\href {\doibase
  10.1103/PhysRevB.106.014301} {\bibfield  {journal} {\bibinfo  {journal}
  {Phys. Rev. B}\ }\textbf {\bibinfo {volume} {106}},\ \bibinfo {pages}
  {014301} (\bibinfo {year} {2022})}\BibitemShut {NoStop}%
\bibitem [{\citenamefont {Jurcevic}\ \emph {et~al.}(2017)\citenamefont
  {Jurcevic}, \citenamefont {Shen}, \citenamefont {Hauke}, \citenamefont
  {Maier}, \citenamefont {Brydges}, \citenamefont {Hempel}, \citenamefont
  {Lanyon}, \citenamefont {Heyl}, \citenamefont {Blatt},\ and\ \citenamefont
  {Roos}}]{Jurcevic2017119}%
  \BibitemOpen
  \bibfield  {author} {\bibinfo {author} {\bibfnamefont {P.}~\bibnamefont
  {Jurcevic}}, \bibinfo {author} {\bibfnamefont {H.}~\bibnamefont {Shen}},
  \bibinfo {author} {\bibfnamefont {P.}~\bibnamefont {Hauke}}, \bibinfo
  {author} {\bibfnamefont {C.}~\bibnamefont {Maier}}, \bibinfo {author}
  {\bibfnamefont {T.}~\bibnamefont {Brydges}}, \bibinfo {author} {\bibfnamefont
  {C.}~\bibnamefont {Hempel}}, \bibinfo {author} {\bibfnamefont {B.~P.}\
  \bibnamefont {Lanyon}}, \bibinfo {author} {\bibfnamefont {M.}~\bibnamefont
  {Heyl}}, \bibinfo {author} {\bibfnamefont {R.}~\bibnamefont {Blatt}}, \ and\
  \bibinfo {author} {\bibfnamefont {C.~F.}\ \bibnamefont {Roos}},\ }\href
  {\doibase 10.1103/PhysRevLett.119.080501} {\bibfield  {journal} {\bibinfo
  {journal} {Phys. Rev. Lett.}\ }\textbf {\bibinfo {volume} {119}},\ \bibinfo
  {pages} {080501} (\bibinfo {year} {2017})}\BibitemShut {NoStop}%
\bibitem [{\citenamefont {Zhang}\ \emph {et~al.}(2017)\citenamefont {Zhang},
  \citenamefont {Pagano}, \citenamefont {Hess}, \citenamefont {Kyprianidis},
  \citenamefont {Ecker}, \citenamefont {Kaplan}, \citenamefont {Gorshkov},
  \citenamefont {Gong},\ and\ \citenamefont {Monroe}}]{Zhang2017551}%
  \BibitemOpen
  \bibfield  {author} {\bibinfo {author} {\bibfnamefont {J.}~\bibnamefont
  {Zhang}}, \bibinfo {author} {\bibfnamefont {G.}~\bibnamefont {Pagano}},
  \bibinfo {author} {\bibfnamefont {P.~W.}\ \bibnamefont {Hess}}, \bibinfo
  {author} {\bibfnamefont {A.}~\bibnamefont {Kyprianidis}}, \bibinfo {author}
  {\bibfnamefont {P.~B.}\ \bibnamefont {Ecker}}, \bibinfo {author}
  {\bibfnamefont {H.}~\bibnamefont {Kaplan}}, \bibinfo {author} {\bibfnamefont
  {A.~V.}\ \bibnamefont {Gorshkov}}, \bibinfo {author} {\bibfnamefont {Z.~X.}\
  \bibnamefont {Gong}}, \ and\ \bibinfo {author} {\bibfnamefont
  {C.}~\bibnamefont {Monroe}},\ }\href {\doibase 10.1038/nature24654}
  {\bibfield  {journal} {\bibinfo  {journal} {Nature}\ }\textbf {\bibinfo
  {volume} {551}},\ \bibinfo {pages} {601} (\bibinfo {year}
  {2017})}\BibitemShut {NoStop}%
\bibitem [{\citenamefont {Guo}\ \emph {et~al.}(2019)\citenamefont {Guo},
  \citenamefont {Yang}, \citenamefont {Zeng}, \citenamefont {Peng},
  \citenamefont {Li}, \citenamefont {Deng}, \citenamefont {Jin}, \citenamefont
  {Chen}, \citenamefont {Zheng},\ and\ \citenamefont {Fan}}]{Guo201911}%
  \BibitemOpen
  \bibfield  {author} {\bibinfo {author} {\bibfnamefont {X.-Y.}\ \bibnamefont
  {Guo}}, \bibinfo {author} {\bibfnamefont {C.}~\bibnamefont {Yang}}, \bibinfo
  {author} {\bibfnamefont {Y.}~\bibnamefont {Zeng}}, \bibinfo {author}
  {\bibfnamefont {Y.}~\bibnamefont {Peng}}, \bibinfo {author} {\bibfnamefont
  {H.-K.}\ \bibnamefont {Li}}, \bibinfo {author} {\bibfnamefont
  {H.}~\bibnamefont {Deng}}, \bibinfo {author} {\bibfnamefont {Y.-R.}\
  \bibnamefont {Jin}}, \bibinfo {author} {\bibfnamefont {S.}~\bibnamefont
  {Chen}}, \bibinfo {author} {\bibfnamefont {D.}~\bibnamefont {Zheng}}, \ and\
  \bibinfo {author} {\bibfnamefont {H.}~\bibnamefont {Fan}},\ }\href {\doibase
  10.1103/PhysRevApplied.11.044080} {\bibfield  {journal} {\bibinfo  {journal}
  {Phys. Rev. Applied}\ }\textbf {\bibinfo {volume} {11}},\ \bibinfo {pages}
  {044080} (\bibinfo {year} {2019})}\BibitemShut {NoStop}%
\bibitem [{\citenamefont {Chen}\ \emph {et~al.}(2020)\citenamefont {Chen},
  \citenamefont {Cui}, \citenamefont {Ai}, \citenamefont {He}, \citenamefont
  {Huang}, \citenamefont {Han}, \citenamefont {Li},\ and\ \citenamefont
  {Guo}}]{Chen2020102}%
  \BibitemOpen
  \bibfield  {author} {\bibinfo {author} {\bibfnamefont {Z.}~\bibnamefont
  {Chen}}, \bibinfo {author} {\bibfnamefont {J.-M.}\ \bibnamefont {Cui}},
  \bibinfo {author} {\bibfnamefont {M.-Z.}\ \bibnamefont {Ai}}, \bibinfo
  {author} {\bibfnamefont {R.}~\bibnamefont {He}}, \bibinfo {author}
  {\bibfnamefont {Y.-F.}\ \bibnamefont {Huang}}, \bibinfo {author}
  {\bibfnamefont {Y.-J.}\ \bibnamefont {Han}}, \bibinfo {author} {\bibfnamefont
  {C.-F.}\ \bibnamefont {Li}}, \ and\ \bibinfo {author} {\bibfnamefont {G.-C.}\
  \bibnamefont {Guo}},\ }\href {\doibase 10.1103/PhysRevA.102.042222}
  {\bibfield  {journal} {\bibinfo  {journal} {Phys. Rev. A}\ }\textbf {\bibinfo
  {volume} {102}},\ \bibinfo {pages} {042222} (\bibinfo {year}
  {2020})}\BibitemShut {NoStop}%
\bibitem [{\citenamefont {Nie}\ \emph {et~al.}(2020)\citenamefont {Nie},
  \citenamefont {Wei}, \citenamefont {Chen}, \citenamefont {Zhang},
  \citenamefont {Zhao}, \citenamefont {Qiu}, \citenamefont {Tian},
  \citenamefont {Ji}, \citenamefont {Xin}, \citenamefont {Lu},\ and\
  \citenamefont {Li}}]{Nie2020124}%
  \BibitemOpen
  \bibfield  {author} {\bibinfo {author} {\bibfnamefont {X.}~\bibnamefont
  {Nie}}, \bibinfo {author} {\bibfnamefont {B.-B.}\ \bibnamefont {Wei}},
  \bibinfo {author} {\bibfnamefont {X.}~\bibnamefont {Chen}}, \bibinfo {author}
  {\bibfnamefont {Z.}~\bibnamefont {Zhang}}, \bibinfo {author} {\bibfnamefont
  {X.}~\bibnamefont {Zhao}}, \bibinfo {author} {\bibfnamefont {C.}~\bibnamefont
  {Qiu}}, \bibinfo {author} {\bibfnamefont {Y.}~\bibnamefont {Tian}}, \bibinfo
  {author} {\bibfnamefont {Y.}~\bibnamefont {Ji}}, \bibinfo {author}
  {\bibfnamefont {T.}~\bibnamefont {Xin}}, \bibinfo {author} {\bibfnamefont
  {D.}~\bibnamefont {Lu}}, \ and\ \bibinfo {author} {\bibfnamefont
  {J.}~\bibnamefont {Li}},\ }\href {\doibase 10.1103/PhysRevLett.124.250601}
  {\bibfield  {journal} {\bibinfo  {journal} {Phys. Rev. Lett.}\ }\textbf
  {\bibinfo {volume} {124}},\ \bibinfo {pages} {250601} (\bibinfo {year}
  {2020})}\BibitemShut {NoStop}%
\bibitem [{\citenamefont {Tian}\ \emph {et~al.}(2020)\citenamefont {Tian},
  \citenamefont {Yang}, \citenamefont {Qiu}, \citenamefont {Liang},
  \citenamefont {Yang}, \citenamefont {Xu},\ and\ \citenamefont
  {Duan}}]{Tian2020124}%
  \BibitemOpen
  \bibfield  {author} {\bibinfo {author} {\bibfnamefont {T.}~\bibnamefont
  {Tian}}, \bibinfo {author} {\bibfnamefont {H.-X.}\ \bibnamefont {Yang}},
  \bibinfo {author} {\bibfnamefont {L.-Y.}\ \bibnamefont {Qiu}}, \bibinfo
  {author} {\bibfnamefont {H.-Y.}\ \bibnamefont {Liang}}, \bibinfo {author}
  {\bibfnamefont {Y.-B.}\ \bibnamefont {Yang}}, \bibinfo {author}
  {\bibfnamefont {Y.}~\bibnamefont {Xu}}, \ and\ \bibinfo {author}
  {\bibfnamefont {L.-M.}\ \bibnamefont {Duan}},\ }\href {\doibase
  10.1103/PhysRevLett.124.043001} {\bibfield  {journal} {\bibinfo  {journal}
  {Phys. Rev. Lett.}\ }\textbf {\bibinfo {volume} {124}},\ \bibinfo {pages}
  {043001} (\bibinfo {year} {2020})}\BibitemShut {NoStop}%
\bibitem [{\citenamefont {Yuzbashyan}\ \emph {et~al.}(2006)\citenamefont
  {Yuzbashyan}, \citenamefont {Tsyplyatyev},\ and\ \citenamefont
  {Altshuler}}]{Yuzbashyan200696}%
  \BibitemOpen
  \bibfield  {author} {\bibinfo {author} {\bibfnamefont {E.~A.}\ \bibnamefont
  {Yuzbashyan}}, \bibinfo {author} {\bibfnamefont {O.}~\bibnamefont
  {Tsyplyatyev}}, \ and\ \bibinfo {author} {\bibfnamefont {B.~L.}\ \bibnamefont
  {Altshuler}},\ }\href {\doibase 10.1103/PhysRevLett.96.097005} {\bibfield
  {journal} {\bibinfo  {journal} {Phys. Rev. Lett.}\ }\textbf {\bibinfo
  {volume} {96}},\ \bibinfo {pages} {097005} (\bibinfo {year}
  {2006})}\BibitemShut {NoStop}%
\bibitem [{\citenamefont {Barmettler}\ \emph {et~al.}(2009)\citenamefont
  {Barmettler}, \citenamefont {Punk}, \citenamefont {Gritsev}, \citenamefont
  {Demler},\ and\ \citenamefont {Altman}}]{Barmettler2009102}%
  \BibitemOpen
  \bibfield  {author} {\bibinfo {author} {\bibfnamefont {P.}~\bibnamefont
  {Barmettler}}, \bibinfo {author} {\bibfnamefont {M.}~\bibnamefont {Punk}},
  \bibinfo {author} {\bibfnamefont {V.}~\bibnamefont {Gritsev}}, \bibinfo
  {author} {\bibfnamefont {E.}~\bibnamefont {Demler}}, \ and\ \bibinfo {author}
  {\bibfnamefont {E.}~\bibnamefont {Altman}},\ }\href {\doibase
  10.1103/PhysRevLett.102.130603} {\bibfield  {journal} {\bibinfo  {journal}
  {Phys. Rev. Lett.}\ }\textbf {\bibinfo {volume} {102}},\ \bibinfo {pages}
  {130603} (\bibinfo {year} {2009})}\BibitemShut {NoStop}%
\bibitem [{\citenamefont {Eckstein}\ \emph {et~al.}(2009)\citenamefont
  {Eckstein}, \citenamefont {Kollar},\ and\ \citenamefont
  {Werner}}]{Eckstein2009103}%
  \BibitemOpen
  \bibfield  {author} {\bibinfo {author} {\bibfnamefont {M.}~\bibnamefont
  {Eckstein}}, \bibinfo {author} {\bibfnamefont {M.}~\bibnamefont {Kollar}}, \
  and\ \bibinfo {author} {\bibfnamefont {P.}~\bibnamefont {Werner}},\ }\href
  {\doibase 10.1103/PhysRevLett.103.056403} {\bibfield  {journal} {\bibinfo
  {journal} {Phys. Rev. Lett.}\ }\textbf {\bibinfo {volume} {103}},\ \bibinfo
  {pages} {056403} (\bibinfo {year} {2009})}\BibitemShut {NoStop}%
\bibitem [{\citenamefont {Sciolla}\ and\ \citenamefont
  {Biroli}(2010)}]{Sciolla2010105}%
  \BibitemOpen
  \bibfield  {author} {\bibinfo {author} {\bibfnamefont {B.}~\bibnamefont
  {Sciolla}}\ and\ \bibinfo {author} {\bibfnamefont {G.}~\bibnamefont
  {Biroli}},\ }\href {\doibase 10.1103/PhysRevLett.105.220401} {\bibfield
  {journal} {\bibinfo  {journal} {Phys. Rev. Lett.}\ }\textbf {\bibinfo
  {volume} {105}},\ \bibinfo {pages} {220401} (\bibinfo {year}
  {2010})}\BibitemShut {NoStop}%
\bibitem [{\citenamefont {Dziarmaga}(2010)}]{Dziarmaga201059}%
  \BibitemOpen
  \bibfield  {author} {\bibinfo {author} {\bibfnamefont {J.}~\bibnamefont
  {Dziarmaga}},\ }\href {\doibase 10.1080/00018732.2010.514702} {\bibfield
  {journal} {\bibinfo  {journal} {Advances in Physics}\ }\textbf {\bibinfo
  {volume} {59}},\ \bibinfo {pages} {1063} (\bibinfo {year}
  {2010})}\BibitemShut {NoStop}%
\bibitem [{\citenamefont {Cao}\ \emph {et~al.}(2022{\natexlab{b}})\citenamefont
  {Cao}, \citenamefont {Zhong},\ and\ \citenamefont {Tong}}]{Cao202236}%
  \BibitemOpen
  \bibfield  {author} {\bibinfo {author} {\bibfnamefont {K.}~\bibnamefont
  {Cao}}, \bibinfo {author} {\bibfnamefont {M.}~\bibnamefont {Zhong}}, \ and\
  \bibinfo {author} {\bibfnamefont {P.}~\bibnamefont {Tong}},\ }\href {\doibase
  10.1088/1751-8121/ac8324} {\bibfield  {journal} {\bibinfo  {journal} {J.
  Phys. A-Math. Theor.}\ }\textbf {\bibinfo {volume} {55}},\ \bibinfo {pages}
  {365001} (\bibinfo {year} {2022}{\natexlab{b}})}\BibitemShut {NoStop}%
\bibitem [{\citenamefont {Tong}\ and\ \citenamefont
  {Zhong}(2001)}]{Tong200191}%
  \BibitemOpen
  \bibfield  {author} {\bibinfo {author} {\bibfnamefont {P.}~\bibnamefont
  {Tong}}\ and\ \bibinfo {author} {\bibfnamefont {M.}~\bibnamefont {Zhong}},\
  }\href {\doibase https://doi.org/10.1016/S0921-4526(01)00546-4} {\bibfield
  {journal} {\bibinfo  {journal} {Physica B}\ }\textbf {\bibinfo {volume}
  {304}},\ \bibinfo {pages} {91} (\bibinfo {year} {2001})}\BibitemShut
  {NoStop}%
\bibitem [{\citenamefont {Tong}\ and\ \citenamefont
  {Zhong}(2002)}]{Tong200265}%
  \BibitemOpen
  \bibfield  {author} {\bibinfo {author} {\bibfnamefont {P.}~\bibnamefont
  {Tong}}\ and\ \bibinfo {author} {\bibfnamefont {M.}~\bibnamefont {Zhong}},\
  }\href {\doibase 10.1103/PhysRevB.65.064421} {\bibfield  {journal} {\bibinfo
  {journal} {Phys. Rev. B}\ }\textbf {\bibinfo {volume} {65}},\ \bibinfo
  {pages} {064421} (\bibinfo {year} {2002})}\BibitemShut {NoStop}%
\bibitem [{\citenamefont {Titvinidze}\ and\ \citenamefont
  {Japaridze}(2003)}]{Titvinidze200332}%
  \BibitemOpen
  \bibfield  {author} {\bibinfo {author} {\bibfnamefont {I.}~\bibnamefont
  {Titvinidze}}\ and\ \bibinfo {author} {\bibfnamefont {G.~I.}\ \bibnamefont
  {Japaridze}},\ }\href {\doibase 10.1140/epjb/e2003-00113-8} {\bibfield
  {journal} {\bibinfo  {journal} {The European Physical Journal B - Condensed
  Matter and Complex Systems}\ }\textbf {\bibinfo {volume} {32}},\ \bibinfo
  {pages} {383} (\bibinfo {year} {2003})}\BibitemShut {NoStop}%
\bibitem [{\citenamefont {Pfeuty}(1979)}]{Pfeuty1979245}%
  \BibitemOpen
  \bibfield  {author} {\bibinfo {author} {\bibfnamefont {P.}~\bibnamefont
  {Pfeuty}},\ }\href {\doibase https://doi.org/10.1016/0375-9601(79)90017-3}
  {\bibfield  {journal} {\bibinfo  {journal} {Phys. Lett. A}\ }\textbf
  {\bibinfo {volume} {72}},\ \bibinfo {pages} {245} (\bibinfo {year}
  {1979})}\BibitemShut {NoStop}%
\bibitem [{\citenamefont {Lieb}\ \emph {et~al.}(1961)\citenamefont {Lieb},
  \citenamefont {Schultz},\ and\ \citenamefont {Mattis}}]{Lieb196116}%
  \BibitemOpen
  \bibfield  {author} {\bibinfo {author} {\bibfnamefont {E.}~\bibnamefont
  {Lieb}}, \bibinfo {author} {\bibfnamefont {T.}~\bibnamefont {Schultz}}, \
  and\ \bibinfo {author} {\bibfnamefont {D.}~\bibnamefont {Mattis}},\ }\href
  {\doibase https://doi.org/10.1016/0003-4916(61)90115-4} {\bibfield  {journal}
  {\bibinfo  {journal} {Annals of Physics}\ }\textbf {\bibinfo {volume} {16}},\
  \bibinfo {pages} {407} (\bibinfo {year} {1961})}\BibitemShut {NoStop}%
\bibitem [{\citenamefont {Suzuki}\ \emph {et~al.}(2013)\citenamefont {Suzuki},
  \citenamefont {Inoue},\ and\ \citenamefont {Chakrabarti}}]{Suzuki2013}%
  \BibitemOpen
  \bibfield  {author} {\bibinfo {author} {\bibfnamefont {S.}~\bibnamefont
  {Suzuki}}, \bibinfo {author} {\bibfnamefont {J.-i.}\ \bibnamefont {Inoue}}, \
  and\ \bibinfo {author} {\bibfnamefont {B.~K.}\ \bibnamefont {Chakrabarti}},\
  }\enquote {\bibinfo {title} {Transverse ising chain (pure system)},}\ in\
  \href {\doibase 10.1007/978-3-642-33039-1_2} {\emph {\bibinfo {booktitle}
  {Quantum Ising Phases and Transitions in Transverse Ising Models}}}\
  (\bibinfo  {publisher} {Springer Berlin Heidelberg},\ \bibinfo {address}
  {Berlin, Heidelberg},\ \bibinfo {year} {2013})\ pp.\ \bibinfo {pages}
  {13--46}\BibitemShut {NoStop}%
\bibitem [{\citenamefont {Zhong}\ and\ \citenamefont
  {Tong}(2011)}]{Zhong201184}%
  \BibitemOpen
  \bibfield  {author} {\bibinfo {author} {\bibfnamefont {M.}~\bibnamefont
  {Zhong}}\ and\ \bibinfo {author} {\bibfnamefont {P.}~\bibnamefont {Tong}},\
  }\href {\doibase 10.1103/PhysRevA.84.052105} {\bibfield  {journal} {\bibinfo
  {journal} {Phys. Rev. A}\ }\textbf {\bibinfo {volume} {84}},\ \bibinfo
  {pages} {052105} (\bibinfo {year} {2011})}\BibitemShut {NoStop}%
\end{thebibliography}%

\end{document}